\documentclass{emulateapj}
\usepackage{lscape}

\newcommand{\balq}{BAL~quasar}

\newcommand{\balqs}{BAL~quasars}
\newcommand{\balQs}{BAL~Quasars}
\newcommand{\uv}{ultraviolet}
\newcommand{\xray}{\hbox{X-ray}}
\newcommand{\asca}{{\emph{ASCA}}}
\newcommand{\chandra}{{\emph{Chandra}}}

\newcommand{\xmm}{\emph{XMM-Newton}}

\newcommand{\rosat}{{\emph{ROSAT}}}

\newcommand{\kms}{\mbox{\,km\,s$^{-1}$}}
\newcommand{\cmsq}{\mbox{\,cm$^{-2}$}}

\newcommand{\flux}{\mbox{\,erg\,cm$^{-2}$\,s$^{-1}$}}
\newcommand{\flam}{\mbox{\,erg\,cm$^{-2}$\,s$^{-1}$\,\AA$^{-1}$}}
\newcommand{\fE}{\mbox{\,photon\,cm$^{-2}$\,s$^{-1}$\,keV$^{-1}$}}
\newcommand{\fnu}{\mbox{\,erg\,cm$^{-2}$\,s$^{-1}$\,Hz$^{-1}$}}
\newcommand{\lumin}{\mbox{\,erg~s$^{-1}$}}
\newcommand{\persec}{\mbox{\,s$^{-1}$}}
\newcommand{\nh}{\mbox{${N}_{\rm H}$}} 
\newcommand{\bj}{$B_{\rm J}$}
\newcommand{\ew}{EW$_{\rm a}$}

\newcommand{\aox}{$\alpha_{\rm ox}$}

\newcommand{\aoxc}{$\alpha_{\rm ox}{\rm (corr)}$}
\newcommand{\aoxl}{$\alpha_{\rm ox}(l_{2500})$}
\newcommand{\fdeep}{$f_{\rm deep}$}
\newcommand{\auv}{$\alpha_{\rm UV}$}
\newcommand{\vmax}{$v_{\rm max}$}
\newcommand{\CIV}{\ion{C}{4}}
\newcommand{\SiIV}{\ion{Si}{4}}
\newcommand{\OVI}{\ion{O}{6}}
\newcommand{\MgII}{\ion{Mg}{2}}
\newcommand{\AlIII}{\ion{Al}{3}}
\newcommand{\HI}{\ion{H}{1}}

\newcommand{\daox}{$\Delta \alpha_{\rm ox}$}
\newcommand{\HR}{{HR}}
\newcommand{\GHR}{$\Gamma_{\rm HR}$}

\newcommand{\pgbal}{{PG~2112$+$059}}

\newcommand{\altcite}{\citealt}

\newcommand{\korista}{KVMW}
\newcommand{\weymann}{WMFH}
\newcommand{\total}{35}
\newcommand{\lbqstotal}{44}
\newcommand{\detected}{27}
\begin{document}
 
 
\shortauthors{Gallagher et al.}
\shorttitle{A \chandra\ Survey of \total\ LBQS BAL Quasars}


\title{An Exploratory \textit{Chandra} Survey of a Well-Defined Sample
  of \total\ Large Bright Quasar Survey Broad Absorption Line Quasars}

\author{S.\ C.~Gallagher,\altaffilmark{1}  
W.\ N.~Brandt,\altaffilmark{2} 
G.\ Chartas,\altaffilmark{2}
R. Priddey,\altaffilmark{3} 
G.\ P.~Garmire,\altaffilmark{2} 
\&
R.\ M. Sambruna\altaffilmark{4}
}  

\altaffiltext{1}{Department of Physics \& Astronomy, University of
  California -- Los Angeles, 430 Portola Plaza, Box 951547, Los
  Angeles CA, 90095--1547, USA; {\em sgall@astro.ucla.edu}}
\altaffiltext{2}{Department of Astronomy and Astrophysics, The
        Pennsylvania State University, University Park, PA 16802, USA;
       {\em niel, chartas, garmire@astro.psu.edu}}
\altaffiltext{3}{Centre for Astrophysics Research, University of
        Hertfordshire, College Lane, Hatfield, Hertfordshire AL10 9AB,
        UK; {\em priddey@star.herts.ac.uk}}
\altaffiltext{4}{NASA's Goddard Space Flight Center, Code 661, Greenbelt, MD 20771, USA; {\em
        rms@milkyway.gsfc.nasa.gov}}

\begin{abstract}
We present 4--7~ks \chandra\ observations of \total\ Broad Absorption
Line (BAL) quasars from the Large Bright Quasar Survey, the largest
sample of sensitive, 0.5--8.0~keV \xray\ observations of this class of
quasars to date.  The limited ranges in both redshift ($z=1.42$--2.90)
and \uv\ luminosity (a factor of $\approx12$) of the sample also make
it relatively uniform.  Of \total\ targets, \detected\ are detected
for a detection fraction of 77\%, and we confirm previous studies that
find \balqs\ to be generally \xray\ weak.  Five of the eight
non-detections are known low-ionization \balqs, confirming reports of
extreme \xray\ weakness in this subset ($\sim10\%$ of optically
selected \balqs).  Those \balqs\ with the hardest \xray\ spectra are
also the \xray\ weakest, consistent with the interpretation that
intrinsic absorption is the primary cause of \xray\ weakness in this
class of quasars as a whole.  Furthermore, the observed trend is not
consistent with simple neutral absorption, supporting findings from
spectroscopic observations of individual targets that \balqs\
typically exhibit complex \xray\ absorption (e.g., partially covering
or ionized absorbers). Assuming normal quasar \xray\ continua and
using the hard-band (observed-frame 2--8~keV) \xray\ flux to `correct'
for the effects of intrinsic absorption at softer energies increases
the relative \xray-to-optical flux ratios to much closer to the range
for normal quasars, further indicating that high-ionization \balqs\
are typically neither intrinsically \xray\ weak nor suffer from
Compton-thick absorption.  In general, we find no evidence for
correlations between X-ray weakness and \uv\ absorption-line
properties, with the exception of a likely correlation between the
maximum outflow velocity of \CIV\ absorption and the magnitude of
\xray\ weakness.  We discuss the implications of our results for
disk-wind models of BAL outflows in quasars.
\end{abstract}

\section{Introduction}
\label{sec:intro}
Since the first surveys with \rosat, Broad Absorption Line (BAL)
quasars have been known to have faint soft \xray\ fluxes compared to
their optical fluxes \citep{KoTuEs1994,GrEtal1995,GrMa1996}.  Given
the extreme absorption evident in the ultraviolet, this soft \xray\
faintness was assumed to result from intrinsic absorption with
\nh$\gtrsim10^{22}$\cmsq\ (assuming neutral gas).  With the
0.5--10~keV response of its detectors, a subsequent \asca\ survey was
able to raise this lower limit for some objects by an order of
magnitude, to \nh$\sim5\times10^{23}$\cmsq\ \citep{GaEtal1999}.  In
all of these studies, the premise of a typical underlying quasar
spectral energy distribution and \xray\ continuum was required.
This reasonable assumption was justified by the remarkable
similarity found between the \uv\ emission-line and continuum
properties of BAL and non-\balqs, suggesting that the ionizing
continua of the two populations are unlikely to be significantly
different (\citealt{WeMoFoHe1991}; hereafter WMFH).  The strong
correlation found by Brandt, Laor, \& Wills (2000; hereafter
BLW)\nocite{BrLaWi2000} for the $z<0.5$ Bright Quasar Survey
quasars \citep[hereafter BQS;][]{pg_ref} between \CIV\ absorption
equivalent width (EW$_{\rm a}$) and faintness in soft \hbox{X-rays}
further supported this assumption.  Subsequently, the observation of
\pgbal\ with \asca\ provided the first direct evidence from \xray\
spectroscopy for intrinsic \xray\ absorption and a normal underlying
\xray\ continuum in a \balq\ \citep{GaEtal2001a}.

\citet{GaBrChGa2002} compiled the results from eight \balqs\ (four with
BALs and four with mini-BALs) with enough counts for independent \xray\
spectroscopic analysis.  They concluded from the spectroscopic
evidence that the intrinsic ultraviolet-to-\xray\ spectral energy
distributions of \balqs\ are consistent with those of typical
radio-quiet quasars.  Furthermore, complex, intrinsic absorption with
$\nh=(0.1-5)\times10^{23}$\cmsq\ was generally evident in the \xray\
spectra.  Subsequently, more spectroscopic observations of individual
\balqs\ with \chandra\ and \xmm\ have generally upheld these results
\citep[e.g.,][]{ChBrGaGa2002,AlGr2003,ChBrGa2003,GrMaEl2003,PageEtal2005}
with only a few exceptions where intrinsic \xray\ faintness cannot be
ruled out with the available low signal-to-noise \xray\ data
\citep[e.g.,][]{MathurEtal2000,SaHa2001}.

Constraining the physical state of the \xray\ absorber, e.g., the
column density, ionization state, covering fraction, and velocity
structure, is the ultimate goal of spectroscopic \xray\ studies of
\balqs.  At present however, only a handful of these generally faint
targets have provided data of sufficient quality for detailed spectral
analysis.  This sample is generally heterogeneous, and thus far there
is no outstanding predictor of observed \hbox{0.5--10.0~keV} flux
based on the \uv\ spectral properties.  In other words, though all
quasars with broad ultraviolet absorption also apparently have
substantial \xray\ absorption, there has been no obvious connection
found between the characteristics of the \uv\ and \xray\ absorbers as
of yet.  The one apparent exception to this finding is that
low-ionization BAL (LoBAL) quasars -- the subset of the \balq\
population with \MgII\ and \AlIII\ BALs in addition to the typical
higher ionization \uv\ BALs such as \CIV\ and \OVI\ -- are notably
\xray\ weaker than normal \balqs\ \citep{GreenEtal2001,GaBrChGa2002}.

In this paper, we present exploratory \chandra\ \citep{chandra_ref}
Advanced CCD Imaging Spectrometer (ACIS; \citealt{acis_ref})
observations of \total\ Large Bright Quasar Survey \balqs.  These
short (4--7~ks) observations are intended to determine the basic
\xray\ fluxes and rough spectral shapes of the sample, and for those
targets that are not detected, sensitive upper limits can be set with
these exposure times.  This strategy was successfully applied in the
survey of ten \balqs\ by \citet{GreenEtal2001} who found that normal
\balqs\ were systematically \xray\ faint. With their high detection
fraction (8 out of 10), composite spectral analysis indicated that
this weakness could be explained by intrinsic absorption with a
partially covering neutral absorber with an average
\nh$\sim7\times10^{22}$\cmsq.  Correcting for this level of absorption
resulted in underlying spectral energy distributions typical of normal
radio-quiet quasars.  The present study applies this fruitful approach
to a much larger and more uniform \balq\ sample.  A primary goal of
this \balq\ \xray\ survey is to investigate the relationship between
\uv\ and \xray\ absorption in luminous quasars in an effort to gain
insight into the mechanism for launching and maintaining energetic
quasar winds.

The sample selection is outlined in \S\ref{sec:sample}; in
\S\ref{sec:obs} the \chandra\ observations and \xray\ data analysis
are described. The possible effects of \uv\ variability on determining
the \xray\ weakness relative to the \uv\ continuum are discussed in
\S\ref{sec:var}, and \S\ref{sec:uvabs} is a description of continuum
and \uv\ absorption-line measurements. The results from \xray\
analysis, a comparison with \uv\ spectroscopic properties, and the
implications for disk-wind models are presented in \S\ref{sec:results}
with the summary and conclusions of the study in \S\ref{sec:conc}.
The cosmology adopted throughout the paper is
$H_0$=70\kms\,Mpc$^{-1}$, $\Omega_{\rm M}$=0.3, and
$\Omega_{\Lambda}$=0.7.

\section{Exploratory \textit{Chandra} BAL Quasar Sample}
\label{sec:sample}

In an effort to increase significantly the number of \balqs\ with
useful \xray\ constraints, we have compiled a large sample of \balqs\
from the Large Bright Quasar Survey \citep*[LBQS;][]{lbqs_ref}.  The
targets are optically bright with \bj\footnote{This blue magnitude is
related to the more standard Johnson blue magnitude:
\bj\,$=B-0.28(B-V)$ \citep{lbqs_ref}.} magnitudes of 16.7--18.8, and
they are drawn from a homogeneous, magnitude-limited quasar survey
that has effective, well-defined, and objectively applied selection
criteria \citep*[e.g.,][]{HeFoCh2001}.  Six known Lo\balqs\ are
included.

This sample of \balqs\ offers the advantages of being much larger and
more homogeneous than those in previous hard-band X-ray surveys.  For
comparison, while the \citet{GreenEtal2001} sample had comparable
sensitivity, the smaller sample size (10 objects) as well as the large
ranges in redshift ($z=0.148$--2.371) and ultraviolet luminosity
(almost a factor of 40) made general conclusions about \balqs\ as a
class more difficult to extract.  Our sample ranges over narrower
spans of both properties with $z=1.42$--2.90 and ultraviolet
luminosities covering only a factor of $\approx12$.  The overall sample
is luminous, with $M_B\approx-26.1$ to $-28.4$ (using the
$K$-corrections for non-BAL quasars of \altcite{CrVi1990}).

To date, our sample includes \total\ of the \lbqstotal\ \balqs\ from
the LBQS with a BAL probability of 1.0 \citep{HewFol2003} and $z>1.4$,
the redshift at which the definitive \CIV\ BAL is shifted into the
wavelength regime accessible with ground-based spectroscopy.  This
sample includes all of the confirmed LBQS \balqs\ identified by
\citet{WeMoFoHe1991} and \citet{GrEtal1995}; several more with more
moderate absorption were subsequently found by \citet{HewFol2003}.
The LBQS \balq\ \xray\ sample has distributions of $z$ and \bj\
consistent with the complete $z>1.4$ LBQS \balq\ sample.  The main
discrepancy between the \xray\ subsample and the complete LBQS \balq\
sample is that the \xray\ subset includes a somewhat larger fraction
of quasars with the largest \uv\ absorption troughs (as indicated by
their BALnicity index values; see below for the exact definition of
this quantity). These \balqs\ were most readily identified by eye from
the LBQS discovery spectra and were therefore the first followed up
with higher quality spectroscopy \citep[e.g.,
WMFH;][]{KoVoMoWe1993}. The quasars in our sample were selected before
the publication of \citet{HewFol2003}.

Our observed \chandra\ sample is listed in Table~\ref{tab:opt} which
includes optical photometry, redshift, the flux density at rest-frame
\hbox{2500\,\AA} ($f_{\rm 2500}$), and Galactic \nh\ for each object.  The
values for $f_{\rm 2500}$ were calculated by averaging the flux
densities in the rest-frame range of $2500\pm25$\,\AA\ from the
optical spectra obtained from Korista et al. (1993; hereafter
\korista)\nocite{KoVoMoWe1993}, the Sloan Digital Sky Survey
\citep[SDSS; ][]{sdss_ref}, or the LBQS \citep{lbqs_ref}.  The spectra
were first normalized using the $g$, $r$, and $i$ SDSS (preferred) or
LBQS \bj\ magnitudes by matching the mean flux density in the spectral
region corresponding to each filter bandpass to the photometry.
Table~\ref{tab:opt} also includes the BALnicity Index (BI). The BI,
roughly equivalent to the \CIV~EW$_{\rm a}$ in velocity units,
specifically includes any contiguous absorption that falls between
3000--25,000~\kms\ blueshifted from the systemic redshift, if the
absorption (at least $10\%$ below the continuum) exceeds 2000~\kms\ in
width (\weymann); the BI values for the entire sample were obtained
from \citet{HewFol2003}. According to WMFH, a \balq\ by definition has
non-zero BI.  Though this definition may omit some quasars with
evident broad absorption, all quasars with non-zero BI are definitely
\balqs.

Only one of the quasars in our sample is known to be formally
radio-loud as defined by radio-to-optical flux ratios, $R^{\star}$ or
$R_{i}\ge1$ (see Table~\ref{tab:opt}), where \mbox{$R^{\star}$ =
$\log(S_{\rm 5 GHz}/S_{\rm 2500\AA})$} \citep{StMoWeFo1992} and
\mbox{$R_{i}$ = $\log(S_{\rm 1.4 GHz}/S_{i})$} \citep{ivezic2002}. The
flux densities at 5~GHz, $S_{\rm 5 GHz}$, are taken from
\citet{StMoWeFo1992}, and $S_{\rm 1.4 GHz}$ values are from the Faint
Images of the Radio Sky at Twenty centimeters survey
\citep[FIRST;][]{first_ref} as reported in \citet{HeFoCh2001}.  The
$R_{i}$ upper limits assume a limiting flux density for the FIRST
survey of 1\,mJy.  The optical flux densities (in mJy), $S_{\rm
2500\AA}$ and $S_{i}$, are taken from this paper (Table~\ref{tab:opt},
column 4) and the SDSS, respectively.  We obtained values or limits
for $S_{\rm 1.4 GHz}$ for three additional quasars from the NRAO VLA
Sky Survey (NVSS) with a limiting flux density of $\sim2.5$\,mJy
\citep{nvss_ref}.  Of these, a point source within $1\farcs4$ of the
optical coordinates of B2211$-$1915 had a measured flux density of
64\,mJy to give $R_{i} = 2.11$.  $S_{i}$ for
the three NVSS objects was calculated by extrapolating from $f_{2500}$
using \auv\ (see \S\ref{sec:uvabs}).

The four quasars without radio constraints are too far south to be
accessible to the VLA.  Based on the radio-loud fraction of the LBQS
determined from a cross-correlation with the FIRST survey
\citep{HeFoCh2001}, we expect at most $\sim12\%$ ($<1$) of these to be
radio-loud, though a more likely value is $\sim6\%$ based on the lower
fraction of radio-detected BAL versus non-\balqs\ \citep{HewFol2003}.

\section{X-ray Observations and Data Analysis}
\label{sec:obs}

Each target was observed at the aimpoint of the back-illuminated S3
CCD of ACIS in faint mode.  The exposure times range from
4.3--7.1\,ks, and they were chosen to obtain detections ($>5$ counts)
even with intrinsic absorption up to $5\times10^{23}$\cmsq\ assuming a
normal underlying radio-quiet quasar spectral energy distribution.
The data were processed using the standard \chandra\ \xray\ Center
(CXC) aspect solution and grade filtering.  In general, we followed
the procedure described in detail in \citet{gall+05}, which we outline
briefly.  Both aperture photometry and the CIAO
3.2\footnote{http://cxc.harvard.edu/ciao} wavelet detection tool {\em
wavdetect} \citep{wavdet_ref} were used in the soft
\hbox{(0.5--2.0~keV)}, hard \hbox{(2.0--8.0~keV)}, and full
\hbox{(0.5--8.0~keV)} bands to determine the measured counts for a
point source in each band.  A 60$\arcsec\times60\arcsec$ image around
the known optical position of each quasar was searched with {\em
wavdetect} at the 5$\times10^{-5}$ false-positive probability
threshold for each band.  This threshold, while looser than those
typically used in X-ray surveys, is appropriate when source
positions are specified {\it a priori}.  Wavelet scale sizes were 1,
1.414, 2, 2.828, 4, and 5.66 pixels. For aperture photometry, the
counts were extracted from circular source cells centered on the
full-band \xray\ centroid positions (or LBQS optical positions for
non-detections) with a 2$\farcs$5 radius. Of the \total\ observed
\balqs, \detected\ were significantly detected in the full band, and
22 (21) were individually detected in the soft (hard) bands.  Our
required optical/\xray\ positional match was $1\arcsec$.  This was met
for all but two targets, B1029$-$0125 and B1240$+$1607, which were
discrepant by $1\farcs4$ and $1\farcs2$, respectively.  A careful
inspection of coincident optical (from the Digitized Sky
Survey\footnote{http://archive.stsci.edu/cgi-bin/dss\_form}) and
\xray\ sources indicated systematic 1$\arcsec$--1$\farcs5$ offsets
between the two consistent with an absolute astrometric offset in the
\chandra\ observations. For the remainder of the detected sources, the
average offset between the X-ray and optical positions was
$0\farcs5$. The background (determined from a source-free annulus
surrounding each target) was in all cases negligible ($<$1 count in
the source region).

From the deep-field X-ray source counts of \citet[][see their
  Fig. 4]{BauerEtal2004}, $\approx300$ sources per deg$^2$ are
  expected at our faintest 0.5--2.0~keV upper limit flux of
  $\sim4\times10^{-15}$\flux.  Within a $1\arcsec$-radius source cell,
  therefore, the probability of a false match is
  $\approx2.5\times10^{-3}$ (per band) for the entire sample of
  \total\ objects.  Therefore, even if all three bands are taken to be
  independent, we do not expect any false matches in this sample.

For each target, to provide a coarse quantitative measure of the
spectral shape we calculated the hardness ratio, defined as
\mbox{\HR\,$=(h-s)/(h+s)$}, where $h$ and $s$ refer to the hard- and soft-band
counts, respectively.  A typical, radio-quiet quasar has a power-law
continuum in the 0.5--10.0~keV band characterized by the photon index,
$\Gamma$, and the 1~keV normalization, $N_{\rm 1 keV}$: $f_{\rm
E}=N_{\rm 1 keV}E^{-\Gamma}$ (\fE).  From spectral fitting of
broad-band \xray\ data, $\Gamma$ is found to average $2.0\pm0.25$ for
radio-quiet quasars \citep[e.g.,][]{GeoEtal2000,ReTu2000}.  The
observation dates, exposure times, count rates, and {\HR}s of the quasars
in the sample are listed in Table~\ref{tab:log}.

To transform the observed \HR\ into $\Gamma$, the \xray\ spectral
modeling tool {\em XSPEC} \citep{xspec_ref} was used as a simulator.
For each observation, the appropriate Galactic column density (see
Table~\ref{tab:opt}), auxiliary response file (arf), and
redistribution matrix file (rmf) were incorporated into the modeling.
The arf and rmf are required to model the response of the
telescope-detector system to incident X-rays; they were generated with
the CIAO tool {\em psextract} for each observation and properly take
into account the time-dependent contamination of the ACIS filter
\citep{acis_contam}.  The detector response to a grid of incident
power-law spectra with varying $\Gamma$ was then simulated.  The
observed \HR\ was compared to the modeled \HR\ values, and the exact
value of the corresponding \GHR\ was determined by interpolating
 the grid values (separated by 0.1 in $\Gamma$).  The errors in
\GHR\ reflect the statistical errors in the \HR.  The modeled
full-band count rate was normalized to the observed full-band count
rate to obtain the power-law normalization, $N_{\rm 1 keV}$.  With
$N_{\rm 1 keV}$ and \GHR, the 0.5--8.0~keV flux, $F_{\rm X}$, and the
flux density at rest-frame 2~keV, $f_{\rm 2 keV}$, were calculated.
The errors quoted for these two values are the Poisson errors
\citep{Gehrels} from the full-band counts.  For the non-detections,
$\Gamma=1.0$ was assumed to calculate the limits on $F_{\rm X}$ and
$f_{\rm 2 keV}$ because these quasars are likely to have hard spectra
from absorption.  Assuming a typical soft quasar continuum with
$\Gamma=2.0$ for quasars that are likely absorbed exaggerates the
measured \xray\ weakness because \hbox{ACIS-S} is significantly more
sensitive from 0.5--2.0~keV.

Lastly, \aox=0.384\,$\log(f_{\rm 2 keV}/f_{\rm 2500})$ \citep{aox_ref}
was calculated, where the factor 0.384 is the inverse of the logarithm
of the ratio of the frequencies at which the flux densities are
measured. We point out that $f_{\rm 2500}$ has not been corrected for
intrinsic reddening of the \uv\ quasar continuum.  Apparently
unreddened quasars exhibit a range of continuum shapes
\citep[e.g.,][]{ric+02}, and so it is not clear how to measure
accurately or correct for reddening in an individual quasar.  However,
in an analysis of 224 SDSS \balqs, \citet{reichar+03} found that
\balqs\ are in general mildly dust-reddened, with an average $E(B-V)$
of 0.023 and 0.077 for high-ionization (HiBAL) and Lo\balqs,
respectively.  At \mbox{2500\,\AA}, correcting for this amount of
extinction (assuming Small Magellanic Cloud reddening and using the
\altcite{fm99} extinction parameterization) increases the \uv\ flux
density by factors of 1.17 and 1.68, corresponding to decreases in
\aox\ of $-0.026$ and $-0.087$ for HiBAL and Lo\balqs, respectively.

The parameter \aox\ is known to correlate with \uv\ luminosity, with
more luminous quasars being relatively \xray\ weaker than less
luminous quasars
\citep[e.g.,][]{AvTa1986,WilkesEtal1994,GrEtal1995,VignaliEtal2003,Strateva+2005}.
To account for this known correlation, we have calculated the
parameter \daox=\aox--\aoxl, where \aoxl\ is the expected \aox\ based
on the \hbox{2500\,\AA} monochromatic luminosity, $l_{\rm 2500}$
\citep[Eq. 6 of][]{Strateva+2005}.  This characteristic actually
compensates somewhat for the effect on \aox\ of dust reddening
mentioned above, in that an underestimate of the \uv\ luminosity at
2500\,\AA\ will lead to a slight overestimate of the expected relative
X-ray luminosity.  Therefore, for an average Hi\balq\ (Lo\balq), the
net effect of dust reddening in \daox\ is $-0.017$ ($-0.056$).  We
note that LBQS quasars are in general quite blue, and so this is
likely at the high end of the average systematic offset for this
particular sample.  In addition, there is an average uncertainty of
$\sim\pm0.1$ in \daox\ from the scatter in the \aox\ vs. $l_{2500}$
quasar distribution. The values for $l_{\rm 2500}$ and the derived
\xray\ properties are presented in Table~\ref{tab:xcalc}.

Based on the \xray\ spectral analysis of \citet{GaBrChGa2002}, the
observed \xray\ continua above rest-frame 5~keV for their
spectroscopic sample of \balqs\ were found to be largely free from the
effects of intrinsic absorption.  In each case, fitting this hard
continuum gave $\Gamma\sim2.0$, consistent with the photon indices
from the final model fits which included complex absorption. If this
situation generally holds true, the intrinsic 2~keV flux density can
be reasonably estimated by using the \xray\ continuum above 5~keV to
normalize a typical quasar power-law spectrum.  Following this
assumption, the counts in the observed-frame 2--8~keV bandpass were
used to normalize a $\Gamma=2.0$ power-law model.  With this
normalization, an ``absorption-corrected'' value for the 2~keV flux
density could be used to estimate an intrinsic \aox, \aoxc.  The final
columns in Table~\ref{tab:xcalc} list \aoxc\ and
$\Delta$\aoxc=\aoxc--\aoxl\ for the 21 hard-band detected \balqs, and
upper limits for the rest.

\section{Ultraviolet Variability and \aox}
\label{sec:var}

The LBQS imaging plates were exposed between \hbox{1975--86}
\citep{lbqs_ref}, at least 15 years prior to the first \chandra\
observations; this observed time difference is reduced by cosmological
time dilation by a factor of \hbox{2.4--3.9}.  Given the propensity of
quasars to exhibit both \uv\ and \xray\ continuum variability, the
mismatch between the epochs of optical and \xray\ observations will
introduce additional uncertainties into the measurement of \aox.  Of
the \total\ \balqs\ presented, 20 have SDSS photometric data included
in Data Release 4.\footnote{http://www.sdss.org/dr4/} These
high-quality data enable an investigation into the effects of \uv\
variability in the calculation of \aox.

To examine the potential effects of \uv\ variability, we directly
compared the flux densities at the central wavelengths, 4627 and
4686\,\AA\ (using 4261 and 3631 Jy for the magnitude zeropoints) of
the LBQS \bj\ and SDSS $g$ filters, respectively
\citep{lbqs_ref,sdss_ref}.  Of the 20 quasars with SDSS data, 45\% (9)
gave $g$ flux densities within 15\% of the \bj\ flux densities,
consistent with the stated LBQS photometric uncertainties
\citep{lbqs_ref}.  The mean and standard deviation of the fractional
difference are $4\%\pm28\%$, indicating that there is no evidence for
a significant systematic offset in the blue photometry from the two
surveys.  Furthermore, while long-term variability increases the
uncertainty in \aox\ by as much as $\sim0.04$, this effect is
typically less than the uncertainty from the \xray\ counting
statistics, and the additional random error will not adversely affect
the calculation of average properties of the sample.  To mitigate the
effects of \uv\ variability as much as possible, we have measured
$f_{2500}$ from spectra that are normalized using available photometry
from the epoch closest to the \chandra\ observations.  We also point
out that \xray\ variability of a factor of a few causing \aox\ changes
of $\approx0.2$ has been seen in at least one luminous \balq\
\citep[e.g., PG~2112$+$059;][]{gall+04}. Though it is unknown how
often this occurs, a few objects in the total sample might be expected
to show such \xray\ variability over timescales of years.

\begin{figure*}[t]
\plotone{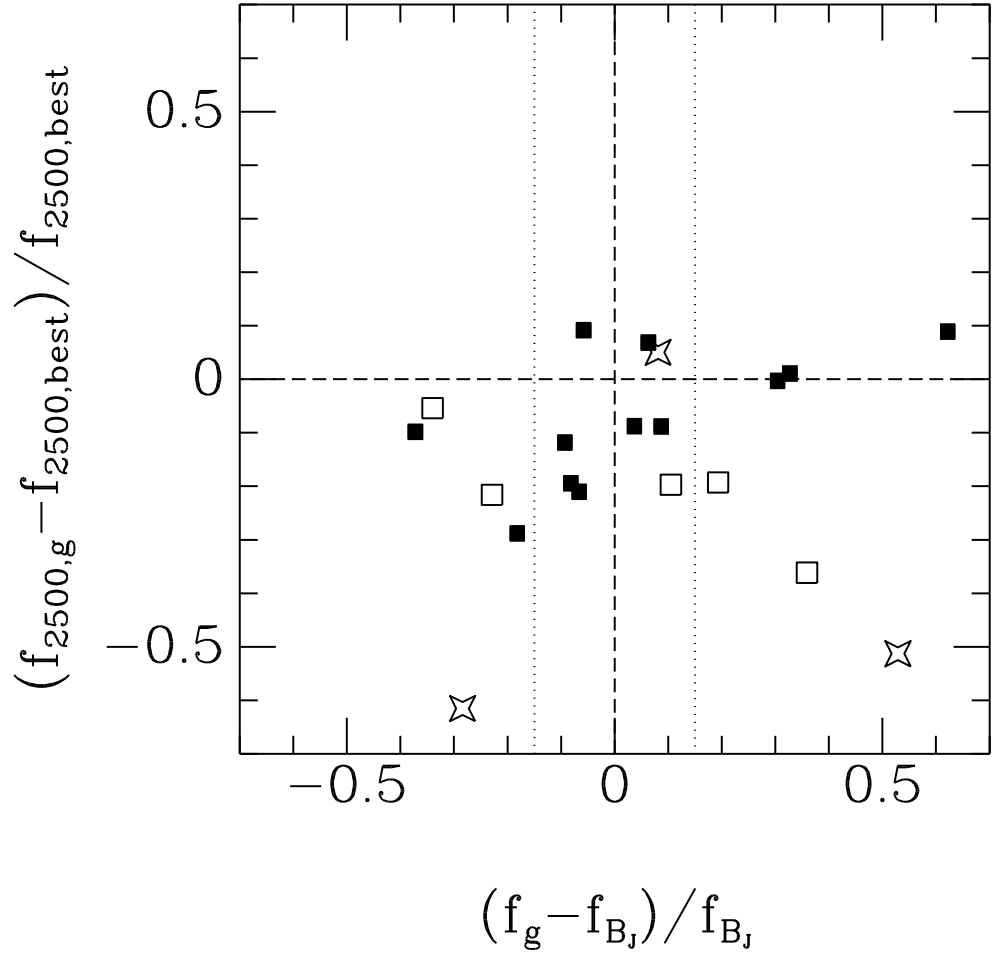}
\caption[]{ Fractional errors in the calculation of $f_{2500}$ from
extrapolation from SDSS $g$ magnitudes to the SDSS photometric band closest in
wavelength to rest-frame 2500\,\AA\ versus fractional uncertainty from
variability between the LBQS and SDSS observations for the 20 LBQS
\balqs\ with SDSS photometry.  Vertical dotted lines indicate the 15\%
LBQS photometric uncertainty \citep{lbqs_ref}.  Open stars mark known
Lo\balqs, filled squares are Hi\balqs, and open squares indicate
\balqs\ of unknown type.
\label{fig:err}
}
\end{figure*}

For reference, we also examined the effects of calculating $f_{2500}$
 by extrapolating from observed-frame blue photometry by assuming a
 typical quasar \uv\ continuum described by a power law with spectral
 index, $\alpha_{\nu}=-0.44$ \citep{VandenBerk2001}.  Specifically, we
 have calculated the rest-frame 2500\,\AA\ flux densities by
 extrapolating from $g$ ($f_{2500,g}$) as well as the SDSS bandpass
 nearest to $2500(1+z)$ ($f_{2500,best}$).  We use $g$ rather than
 \bj\ as data from the same epoch eliminate the effect of variability.
 The mean and standard deviation of the fractional difference between
 $f_{2500,g}$ and $f_{2500,best}$ is $27\%\pm55\%$.  This would give
 on average a systematic underestimate of $f_{2500}$ resulting in a
 mean \aox\ underestimate of $\sim0.04$.  This result is expected
 based on the conclusion of \citet{reichar+03} that \balqs\ typically
 show more continuum reddening than typical quasars with Lo\balqs\
 showing the most reddening \citep{SpFo1992}.  Our measurements of
 $f_{2500}$ were done using spectra, and so this problem does not affect
 our results.  The fractional differences from both the \uv\ variability
 and extrapolation analyses are plotted in Figure~\ref{fig:err}.

\section{Ultraviolet Continuum and Absorption-Line Measurements}
\label{sec:uvabs}

In order to characterize the \uv\ continua and \CIV\ absorption lines
of the LBQS \xray\ sample, we measured several properties from the
available spectroscopy from \korista, SDSS, or the LBQS discovery
spectra (listed in order of preference based on spectral quality). The
first measured parameter was the power-law continuum spectral index,
\auv, where $f_{\lambda}\propto\lambda^{\alpha_{\rm UV}}$, derived
from fitting a power-law to the continuum-dominated rest-frame
spectral regions of \hbox{1700--1800}, \hbox{1950--2050},
\hbox{2150--2250}, and \hbox{2950--3100\,\AA}, as available.  These
values are listed in Table~\ref{tab:opt}.

The BI values for the complete sample (listed in Table~\ref{tab:opt})
were already available.  However, for comparison with other quasar
samples, the \CIV~EW$_{\rm a}$ (in \AA) is also useful.  Though BI
and \CIV~EW$_{\rm a}$ are highly correlated, BI by definition does
not include low-velocity absorption, and so a simple unit conversion
from BI measured in \kms\ to \AA\ will underestimate the corresponding
\CIV~EW$_{\rm a}$.  For the quasars presented in \korista, we were
able to use their continuum-divided spectra (S. Morris, 2005, priv.
comm.).  For the rest, we followed the prescription in \korista\ by 
fitting a low-order polynomial to smooth regions of the continuum to
normalize the spectrum.  Next, we fit a gaussian to the red-wing of
the \CIV\ emission-line, and again normalized the spectrum; this was
necessary to include low-velocity absorption in the \CIV~EW$_{\rm
a}$.  Spectral regions covering the blueshifted outflow velocities of
0--25,000 \kms\ with relative flux densities $<1.0$ were considered
absorbed; this specification gave values that agreed best with the
values for the 23 \balqs\ presented in \citet[][F. Hamann, 2001,
priv. comm.]{HaKoMo1993}.  Our measured \CIV~EW$_{\rm a}$ values are
listed in Table~\ref{tab:opt}.

Based on the correlation of the maximum outflow velocity, \vmax, of
\CIV\ absorption with $M_V$ found by \citet{LaoBra2002} for the \xray\
weak BQS quasars, the quantity \vmax\ is also of interest.  Though
many quasars in the LBQS \xray\ sample have \vmax\ values available in
the literature \citep{LaHu2004}, we redid these measurements for the
entire sample using electronic spectra to obtain a uniform and
consistently measured quantity.  As with the \CIV~EW$_{\rm a}$, the
largest uncertainty in determining \vmax\ is setting the continuum
properly, particularly shortward of the \CIV\ emission line.  Spectral
regions covering the blueshifted outflow velocities of 0--25,000 \kms\
from \CIV$\lambda 1549.06$ where the flux dropped to $\le90\%$ of the
continuum level were considered absorbed.  Though BAL outflow
velocities can be greater than 25,000\kms, above this value \SiIV\
absorption can confuse the measurement. We formally define the
quantity \vmax\ to be the maximum velocity where the absorption was at
least 2000\kms\ wide.  This last requirement was necessary given the
variable spectral quality (both in resolution and signal-to-noise
ratio) in this region.  Those quasars with absorption that starts at
outflow velocities $<25,000$\kms\ and clearly continues above this
value are given lower limits, \vmax$>25,000$\kms.

Finally, we also measured \fdeep, the fraction of the absorption
between 0 and 25,000 \kms\ that is deeper than $50\%$ of the
continuum level.  This last quantity was chosen to investigate the
possibility that quasars with shallow BALs have lower \xray\
absorption column densities, as found for PG~2112$+$059
\citep{GaEtal2001a} and CSO~755 \citep{ohad+05}.  We confirmed by
visual inspection that those \balqs\ with qualitatively shallow
troughs had values of \fdeep$\lesssim0.1$.  The quantities \vmax\ and
\fdeep\ are listed in Table~\ref{tab:opt}.

\section{Results and Discussion}
\label{sec:results}

\subsection{General X-ray Properties of \balQs}

\begin{figure*}[t]
\plotone{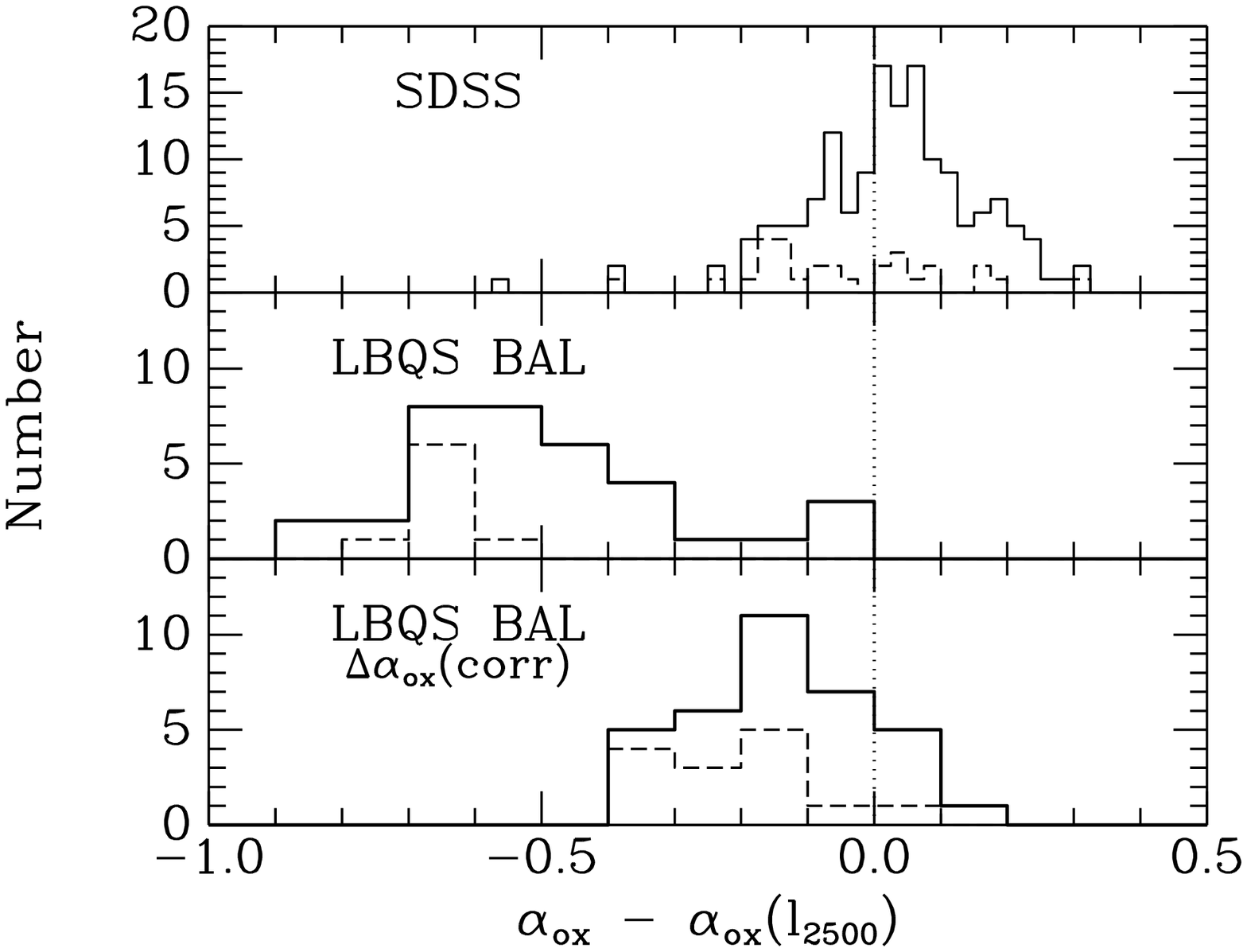}
\caption[]{
\label{fig:hist}
Histograms of the distributions of observed \daox=\aox--\aoxl\ for
the SDSS quasars in the survey of \citet{Strateva+2005} ({\it Top}),
and the LBQS \xray\ \balq\ sample ({\it Middle}). {\it Bottom panel:} The
distribution of $\Delta$\aoxc=\aoxc--\aoxl\ values for the LBQS \xray\
sample (where \aoxc\ is calculated assuming  $\Gamma$=2.0 and normalizing
the \xray\ continuum using the 2--8~keV count rate).  
For all three panels, dashed histograms indicate upper limits, and the
dotted vertical lines mark \daox\,$=0$.
}
\end{figure*}

For the entire sample, we were able to measure or place sensitive
upper limits on \aox, a standard indicator of the relative \xray\
versus \uv\ power.  Our values of \aox\ range from $<$--2.48 to
$-1.65$ with a median value of $-2.20$.  This is an enormous range in
X-ray-to-optical flux ratio of $\sim145$. In Figure~\ref{fig:hist}, we
show the distribution of \daox=\aox--\aoxl\ (to account for the
luminosity dependence of \aox) for this sample.  For comparison, the
distribution of \daox\ values from the SDSS/\rosat\ survey of
\citet{Strateva+2005} is also plotted.  The median quasar in the LBQS
\balq\ sample has an offset of $-0.52$ in \daox\ from the peak of the
typical quasar distribution.  This indicates that the median \balq\ is
$\approx22$ times fainter at rest-frame 2~keV than expected based on
its 2500\,\AA\ \uv\ luminosity.  The distribution of
$\Delta$\aoxc=\aoxc--\aoxl, where \aoxc\ is calculated by assuming
$\Gamma=2$ and using the hard-band count rate to normalize the \xray\
continuum (see \S\ref{sec:obs}), shifts the \balq\ population much
closer to that of normal quasars, with a median value of
\mbox{$\Delta$\aoxc=--0.14}.  The large number of upper limits on
\aoxc\ (14 of \total) makes the distribution of $\Delta$\aoxc\ less
well defined than that of \daox.  However, it is nonetheless clear
that the distribution of $\Delta$\aoxc\ does not agree statistically
with the \citet{Strateva+2005} sample (see Table~\ref{tab:aox}). It is
likely that additional absorption is not being properly accounted for
with this simple correction. The quartile divisions, medians, and
means of the \aox, \daox, and $\Delta$\aoxc\ distributions for the
entire sample and the subset of known Hi\balqs\ are listed in
Table~\ref{tab:aox}.

Notably, of the eight \balqs\ with only upper limits to \aox, five of
these are Lo\balqs; all six Lo\balqs\ have \daox$<-0.52$, the median
of the \balqs{'} distribution.  The only detected known Lo\balq\ in the
sample, B1331$-$0108, has \daox$=-0.67$.  This result strengthens the
conclusion of \citet{GreenEtal2001} that Lo\balqs\ are significantly
\xray\ weaker than normal \balqs.

To obtain tighter constraints on the average \xray\ flux of these
Lo\balqs, we stacked the full, soft, and hard-band images of the five
undetected Lo\balqs\ aligned by the optical quasar positions, and
performed aperture photometry as described in $\S$\ref{sec:obs} on
each composite image.  Stacking analysis is a robust technique that
has been employed successfully to constrain the average X-ray
properties of extremely X-ray faint populations \citep[][see their
Appendix A]{GrEtal1995,stacking_ref}.  The combined exposure time
is \hbox{28.2\,ks}.  In the source cell of the full/soft/hard band
images, 4/3/1 counts were obtained where 1.5/0.4/1.1 counts were
expected from the background.  According to Poisson statistics, this
indicates a significant detection of the Lo\balqs\ in the stacked soft
band image; the Poisson probability of obtaining $\ge3$ counts when
0.4 is expected is $7.9\times10^{-3}$.  This detection yields an
average $F_{\rm X}\sim1\times10^{-15}$\flux\ and \GHR$>0.7$.  Given
the larger effective area in the soft band, this does not provide an
interesting constraint on the \xray\ spectral shape.  Taking the mean
$f_{\rm 2500}$ for the five objects, the average \daox\ is $-0.9$,
$\sim220$ times fainter than expected based on the \uv\ luminosities.
This value for the average \daox\ may result from Compton-thick
absorption as has been spectroscopically demonstrated for the Lo\balq\
Mrk~231 \citep{braito+04}, which has an intrinsic-to-observed
2--10\,keV luminosity ratio of $\sim100$.

While \daox\ is a sensitive indicator of relative \xray\ strength or
weakness, the \xray\ data provide additional information on the shape
of the \xray\ continuum from the distribution of \GHR, which can only
be measured for the \detected\ detected \balqs.  For those objects
with only soft-band detections, \GHR\ is a lower limit; conversely
with only a hard-band detection \GHR\ is an upper limit.

\begin{figure*}
\epsscale{.9}\plotone{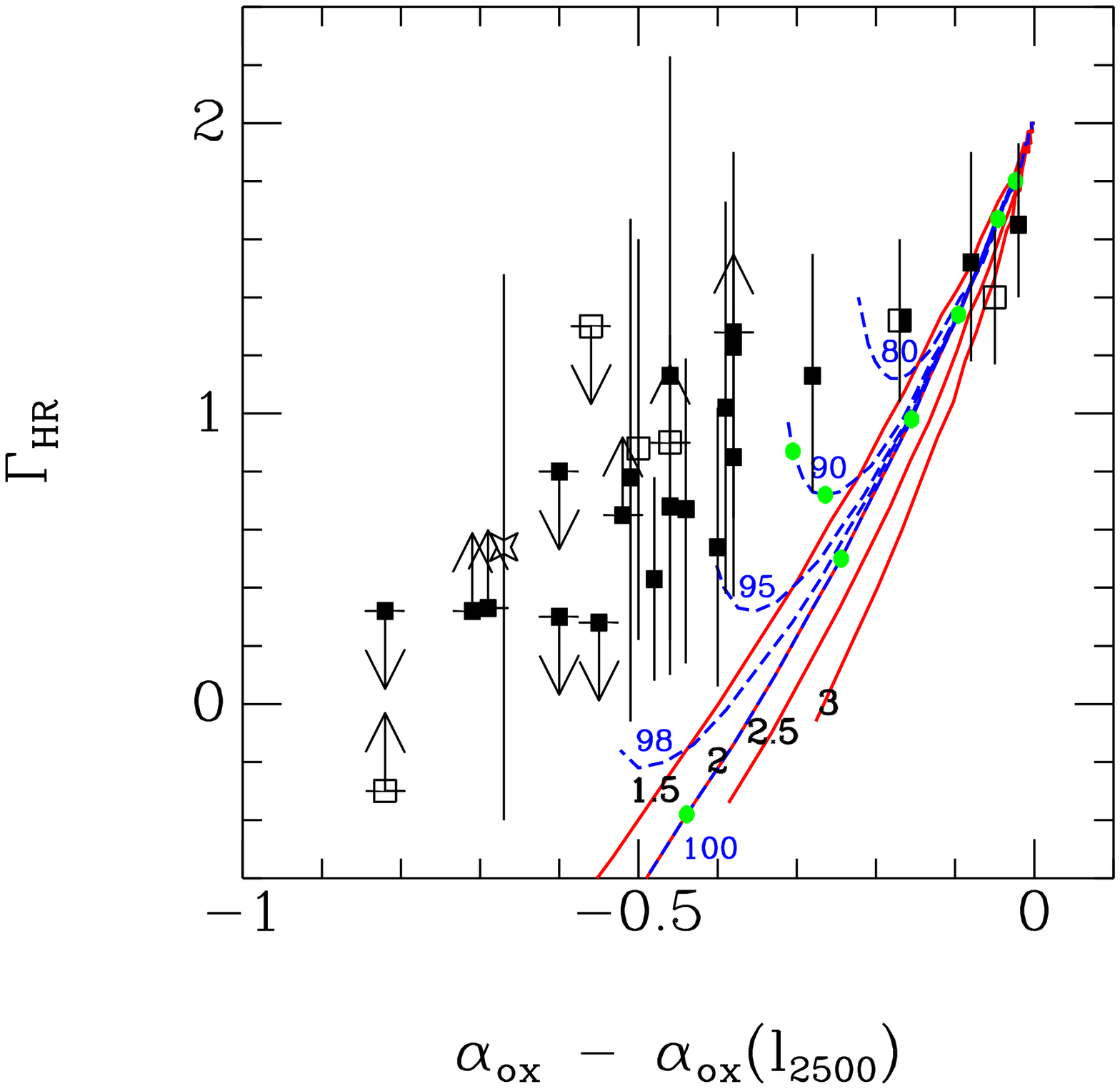}\epsscale{1}
\caption[]{ Plot of \GHR\ versus \daox=\aox--\aoxl\ for the detected
\balqs\ in this sample; the symbols are as in Figure~\ref{fig:err}
with the radio-loud \balq\ marked with a half-filled square.  Solid,
red lines mark the tracks of \GHR\ versus \daox\ for a neutral
absorber assuming $\Gamma=2.0$ where no \xray\ absorption corresponds
to \daox=0, and the absorber is assumed to only affect the \xray\
flux. The column density increases to the left and down along the
tracks. Black labels indicate the redshift of the intrinsic absorber.
The locations of the data points to the left of the tracks are
consistent with complex absorbers, e.g., ionized or partially
covering, that do not block all of the soft photons. Blue, dashed
tracks show the effects of partially covering neutral absorption at
$z=2$; $f_{\rm cov}$, the percentage of continuum covered by the
absorber, is labeled in blue. The tracks curve to larger values of
\GHR\ when the unabsorbed continuum (1--$f_{\rm cov}$) begins
dominating the observed \xray\ flux.  For reference, green circles mark
(0.1,0.2,0.5,1,2,5)$\times10^{23}$\cmsq\ on the red $z=2$ track and
(5,10)$\times10^{23}$\cmsq\ on the blue 90\% track.
\label{fig:hr}
}
\end{figure*}

While typical quasars have rest-frame 2--10~keV photon indices of
$\Gamma\approx2.0$, a quasar exhibiting intrinsic absorption will have
an apparently harder \xray\ spectrum due to the lack of low-energy
\hbox{X-rays}.  In this case, intrinsic absorption would lead to a
smaller value of \GHR; the lack of soft \xray\ photons will also
depress \daox.  As seen in Figure~\ref{fig:hr}, a plot of \GHR\ versus
\daox\ for the sample of detected \balqs, those quasars with the most
negative values of \daox\ tend to have the lowest values of \GHR.
That is, the \xray\ weakest objects also tend to have the \xray\
hardest spectra. To test if this apparent trend represents a
significant correlation, we used the ASURV Rev. 1.2 \citep{LaIsFe1992}
software package to perform a non-parametric, bivariate statistical
test on the data.  ASURV implements the methods described in
\citet*{IsFeNe1986} and was designed to handle censored data sets,
such as \GHR\ (with both upper and lower limits).  For this
application, the generalized Kendall's $\tau$ \citep[][Chapter
14]{Recipes} statistic was calculated.  From the distribution of this
statistic expected from unrelated data, the probability that two
variables are correlated can then be estimated; the results of this
analysis are presented in Table~\ref{tab:stats}.  From the Kendall's
$\tau$ value of 4.055, the probability of no correlation is 0.01$\%$.
Excluding the radio-loud \balq\ B2211$-$1915 did not significantly
alter the result.  We therefore consider \GHR\ and \daox\ to be
significantly correlated.  Testing \GHR\ vs. \aox\ also indicates a
significant correlation independent of \aoxl.

There is no obvious reason why \GHR\ and \daox\ would be correlated if
\balqs\ were typically intrinsically \xray\ weak; intrinsic absorption
provides a reasonable common cause for the correlation.  However, the
tracks of \GHR\ versus \daox\ for intrinsic, neutral absorption shown
in Figure~\ref{fig:hr} lie almost entirely to the right of the actual
data.  This discrepancy does not result from \uv\ reddening as any
corrections to \daox\ to account for reddening would push the observed
\balqs' \daox\ distribution further to the left.  \xray\ spectroscopy
of \balqs\ has shown that the intrinsic absorption in \balqs\ is
typically complex, perhaps due to partial covering and/or ionization,
but with the present spectral resolution and signal-to-noise ratio,
the specific nature of that complexity is poorly constrained.
Empirically, this complexity manifests itself as additional flux at
low X-ray energies from what would be expected from a completely
neutral absorber.  This is also what the data points indicate in
Figure~\ref{fig:hr}.  The dashed, blue tracks in the figure show the
effect of a partially covering neutral absorber in the \GHR-\daox\
plane; though this type of absorption is more consistent with the data
points, they are still more \xray\ weak than would be expected.  A
possible explanation for this is Compton scattering, which is not
included in the {\em XSPEC} absorber models (e.g., {\em wabs} and {\em
zpcfabs}).  Compton scattering is likely to have a measurable effect
for \nh$>10^{22}$\cmsq\ and may account for the depressed \daox\
values.  Unfortunately, satisfactory models for high column density,
ionized absorbers are not available in {\em XSPEC}.

\subsection{Comparison of X-ray and Ultraviolet Properties}
\label{sec:compuv_x}

A primary goal of this survey is to investigate the relationship
between the \uv\ absorber and \xray\ properties to understand better
the nature of the BAL wind.  To characterize quantitatively the \uv\
BALs, we use four parameters, BI, the Detachment Index (DI), \vmax,
and \fdeep; all four were measured for the \CIV\ BAL.  The second
absorption parameter, DI, is defined as a dimensionless measure of the
velocity of the onset of absorption normalized to the width of the
emission line (\weymann).  The values for DI were compiled by
\citet{LaHu2004}; 27 \balqs\ have DI data.  The specific definitions
and measurements of BI, \vmax, and \fdeep\ are described in
\S\ref{sec:uvabs} and listed in Table~\ref{tab:opt}.

\begin{figure*}
\plottwo{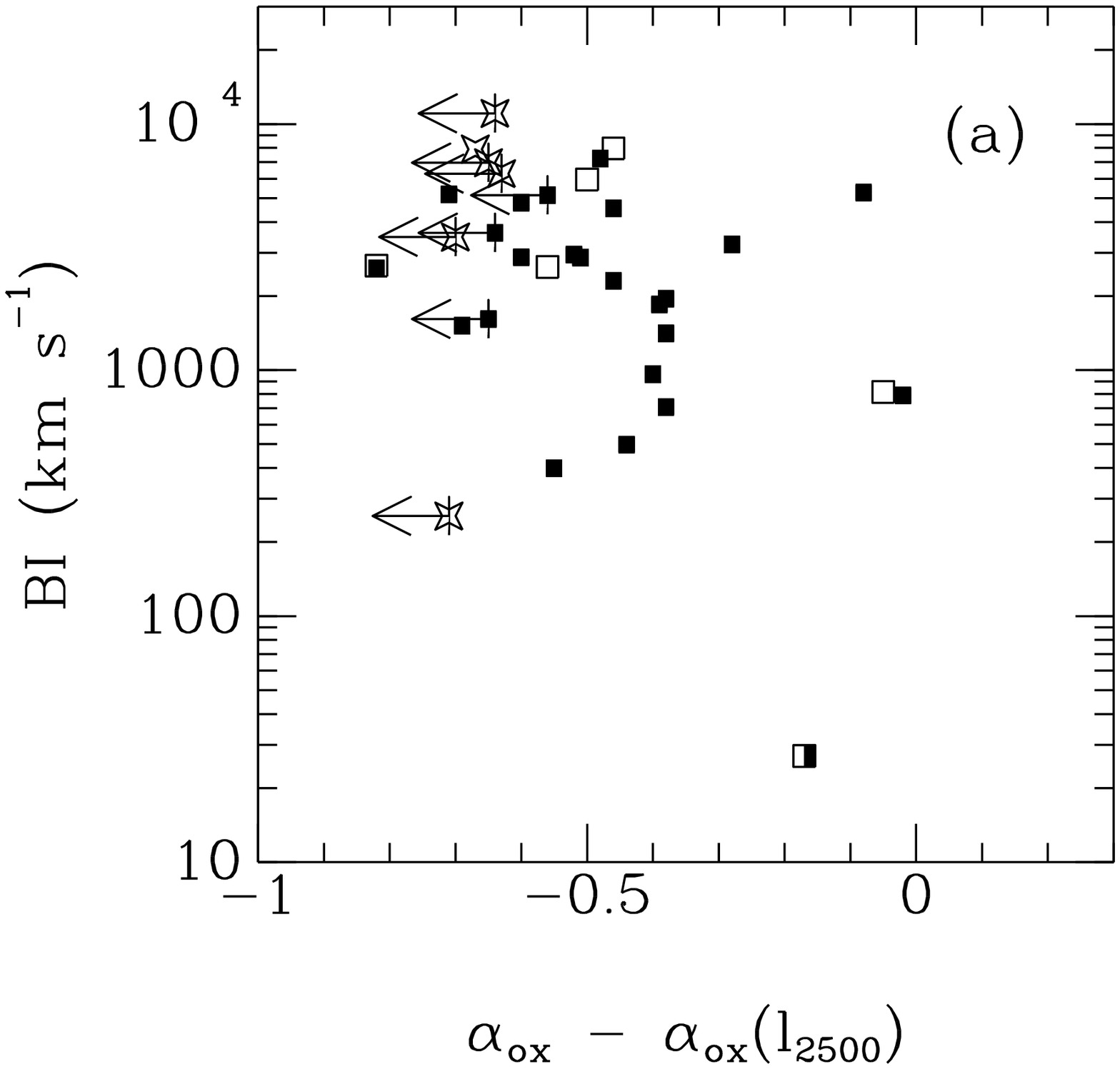}{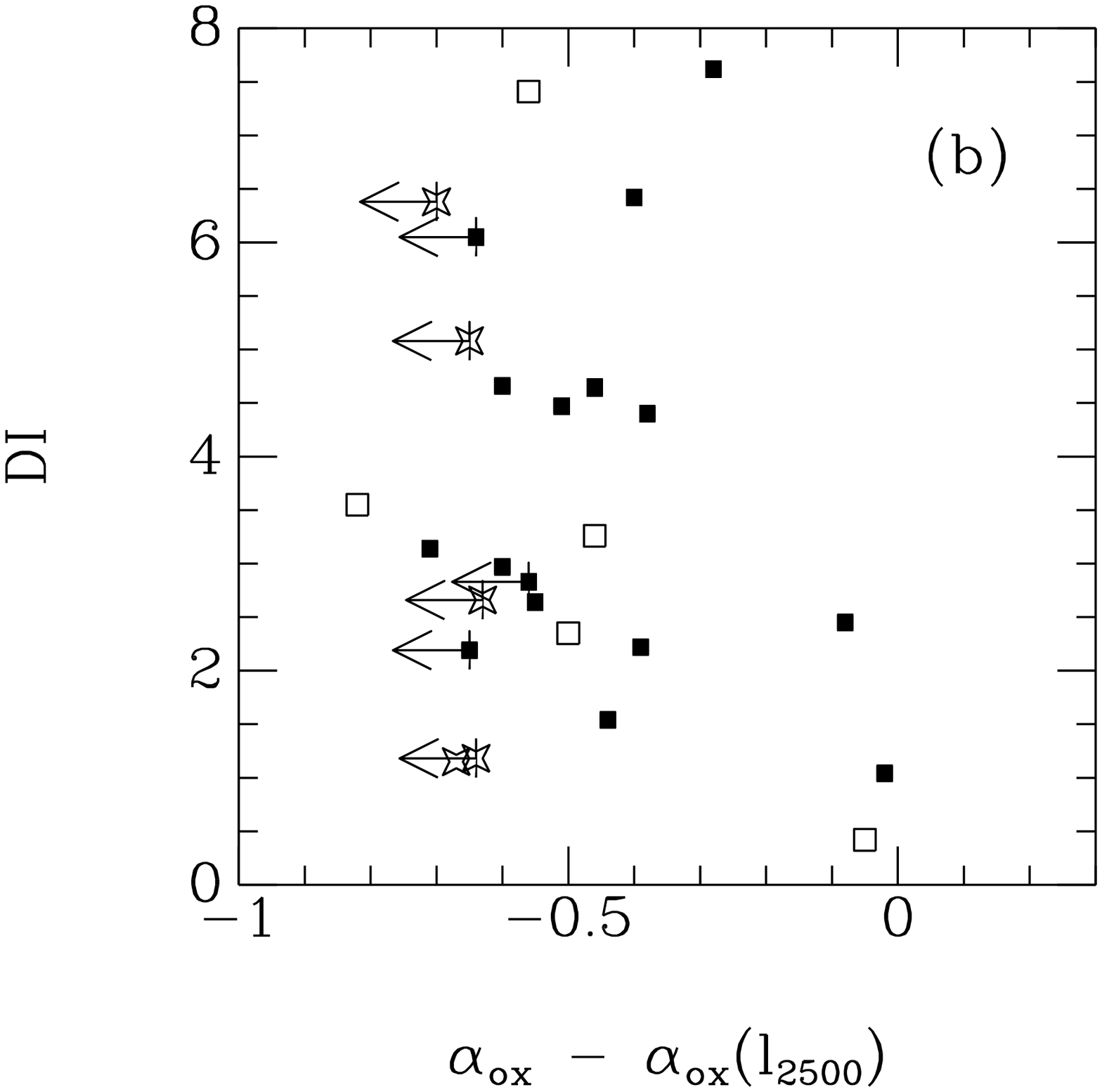}
\plottwo{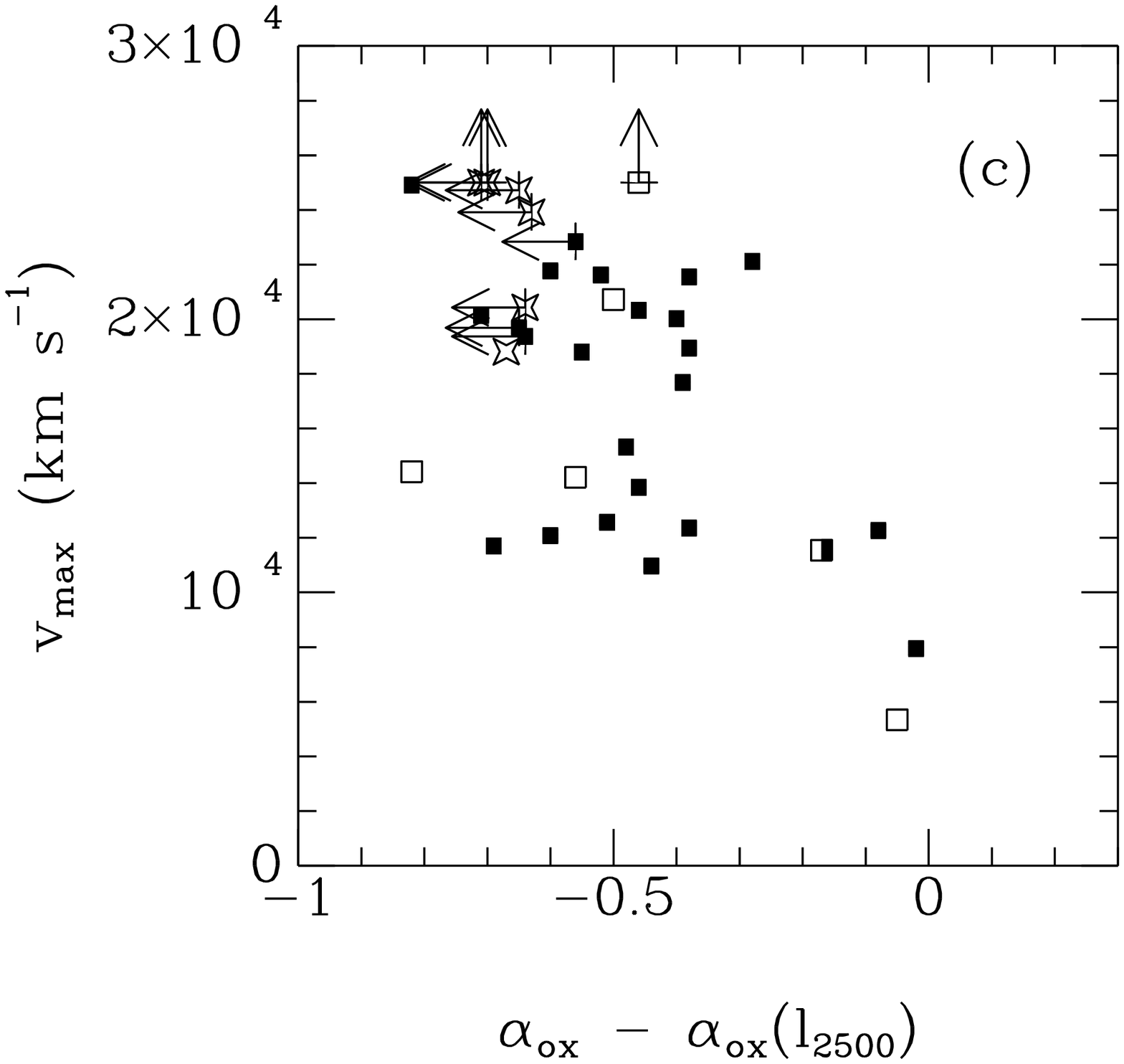}{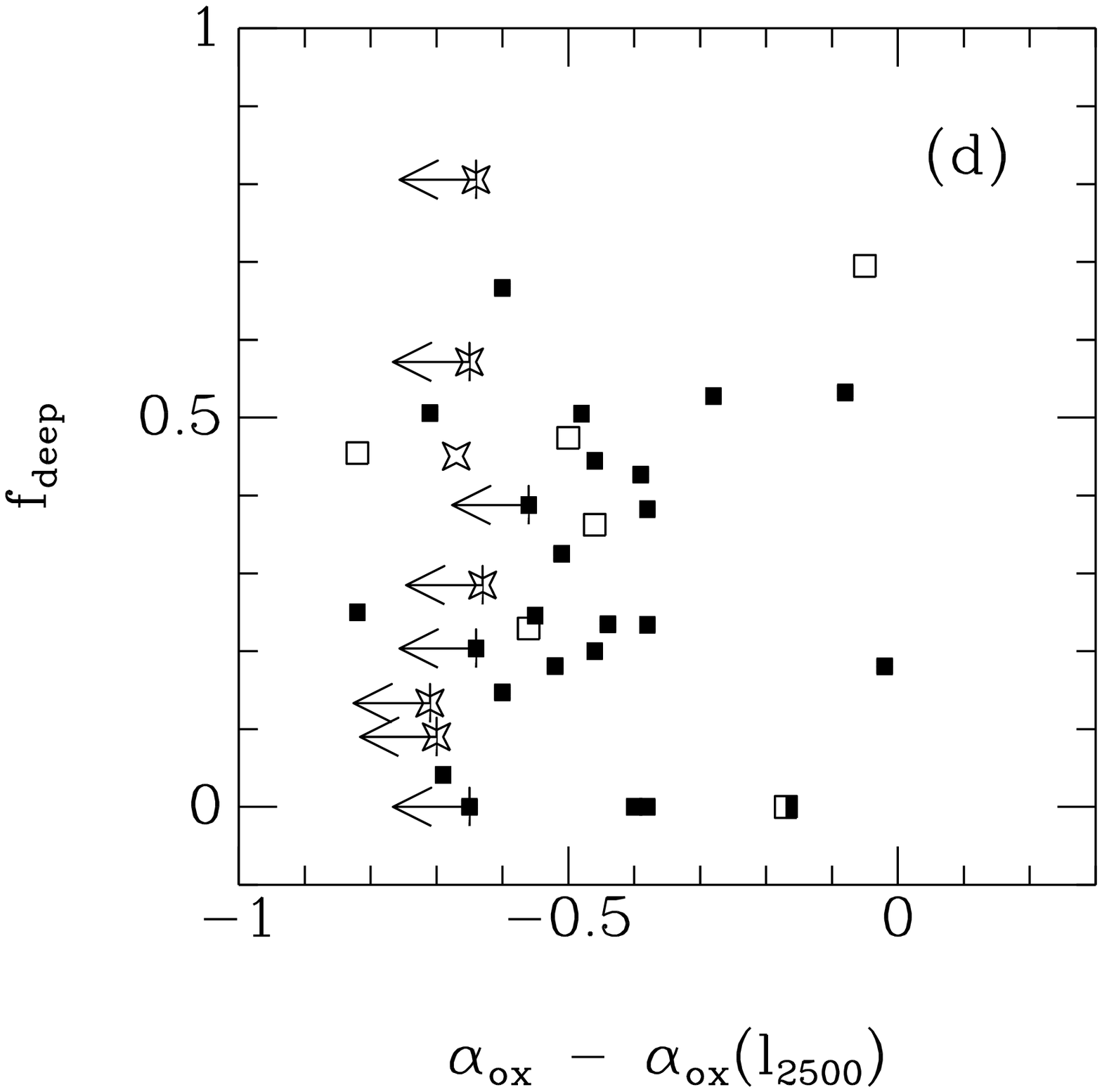}
\caption[Absorption-line parameters versus \daox]{
\CIV\ absorption-line parameters, {\bf (a)} BALnicity
Index (BI), {\bf (b)} Detachment Index (DI), {\bf (c)} maximum outflow
velocity of absorption (\vmax), and {\bf (d)} fraction of the
absorption trough deeper than 50$\%$ of the continuum (\fdeep) are
plotted against \hbox{\daox=\aox--\aoxl}.  Symbols are the same as in
Figure~\ref{fig:hr}.
\label{fig:abs}
}
\end{figure*}

The parameter \daox\ was tested against each \uv\ absorption-line
property using two bivariate non-parametric statistics, generalized
Kendall's $\tau$ and Spearman's $\rho$ \citep[][Chapter 14]{Recipes},
to search for correlations.  We use \daox\ rather than \GHR\ as the
\xray\ parameter of interest because \daox\ is much less sensitive to
Poisson noise and is available for all objects in the sample; only
detected objects have values of \GHR.  We consider two parameters to
be significantly correlated if at least one test gives a probability
that they are not correlated $<0.01$.  In addition to the full sample
of \balqs, the tests were also performed separately for the 24 quasars
known to have only high-ionization BALs (indicated with `Hi' in
Table~\ref{tab:opt}).  We also checked that the presence of the
radio-loud \balq\ B2211$-$1915 in the sample did not significantly
affect any test results; it did not.  The results from these analyses
are presented in Table~\ref{tab:stats}, and each absorption-line
parameter is plotted against \daox\ in Figure~\ref{fig:abs}.

For the first two parameters, BI and DI, there is no significant
evidence for correlations with \daox.  Specifically, the \xray\
weakest \balqs\ and the \xray\ brightest \balqs\ span the range of BI
and DI.  One plausible and straightforward explanation for the lack of
correlation between BI and \daox\ is that the \uv\ and \xray\
absorption are not occurring in the same gas; this has already been
suggested by absorption studies of individual \balqs\ with
spectroscopic \xray\ data.  This first result is perhaps surprising
given that BLW found a highly statistically significant correlation
between \CIV~EW$_{\rm a}$ and \aox\ for BQS quasars.\footnote{We
choose to use BI instead of \CIV~EW$_{\rm a}$ for this analysis as BI
values have been uniformly measured for the entire sample and are less
sensitive to assumptions about the shape of the \CIV\ emission line.}
We compare the BQS with our sample results in more detail below in
\S\ref{sec:bqs}.

The lack of correlation between DI and \daox\ is surprising based on
prior expectations from disk-wind models.  As stated explicitly by
\citet[][\S4]{Goodrich1997}, one would predict from the geometry of
the equatorial disk-wind models of \citet{KoKa1994} and
\citet[][hereafter MCGV]{MuChGrVo1995} that the \balqs\ viewed through
lines of sight closest to the plane of the accretion disk would show
low-velocity \uv\ troughs (and thus have low DI values) as well as the
largest amounts of overall absorption.  We discuss the implications of
this result in more detail in \S\ref{sec:diskwind}.

The third absorption-line parameter, \vmax, does show a significant
correlation with \daox, with probabilities that the properties are not
correlated of 0.003 from both the Kendall and Spearman tests.
(Excluding the radio-loud \balq\ gives probabilities of 0.009 and
0.007 for the Kendall and Spearman tests, respectively.)  Tests of
\vmax\ versus \aox\ (without any dependence on the
\altcite{Strateva+2005} correlation) is similarly signficant.  This
intriguing result indicates that the most \xray\ absorbed \balqs\ (as
indicated by their \xray\ weakness) are more likely to have
high-velocity outflows viewed along the line of sight.  In the context
of radiatively driven winds, this points toward large radial
acceleration of the outflow requiring a very thick \xray\ absorber.
This is most evident in the extreme objects with lower \vmax\ limits
and upper \daox\ limits; both this subset and the low \vmax, \xray\
bright \balqs\ contribute to the significance of the correlation.  At
present, with only four data points at the low \vmax, \xray-bright
end, further testing with larger samples is required to determine if
this correlation holds up with a more uniform distribution of \vmax.
Such studies should also take care to maintain a uniform \uv\
luminosity given the correlation between \vmax\ and \uv\ luminosity
found by \citet{LaoBra2002} for their BQS sample.  Even with the
current sample (as seen in Fig.~\ref{fig:abs}c), the Lo\balqs\ are
preferentially found with the most extreme values of both \daox\ and
\vmax.  One plausible explanation is that both the highest velocity
acceleration and the survival of low-ionization gas in the outflow
require X-ray absorption near the Compton-thick limit.

The final absorption-line property, \fdeep, is not correlated with
\daox.  The six \balqs\ with the shallowest BALs, as identified by
visual inspection and indicated with values of \fdeep$<0.1$, span the
range of \daox.  Though there are some BAL quasars in addition to
PG~2112$+$059 and CSO~755 with shallow troughs that are relatively
\xray\ bright, \fdeep\ is not generally a predictor of the level of
\xray\ weakness.  That is, not all \balqs\ with shallow \uv\
absorption troughs are relatively \xray\ bright, and not all \xray\
bright \balqs\ have shallow troughs.

The \uv\ continuum power-law slope, \auv, shows no evidence of a
correlation with \daox: both red and blue \balqs\ can exhibit extreme
\xray\ weakness.

\subsection{The X-ray Brightest \balQs}
\label{sec:joint}
Even before renormalizing the \xray\ continuum using the
observed-frame 2--8~keV counts (rest-frame $\gtrsim5$~keV), not all of
the \balqs\ are observed to have \daox\ values that would group them
as \xray\ weak, i.e., with \mbox{\daox\,$<-0.2$}.  As seen in
Figure~\ref{fig:hr}, this subset of \balqs\ also has the largest \GHR\
values in the sample with \GHR\,$\sim1.5$.  While the calculated
\xray\ fluxes do suffer from large errors due to the generally low
count rates, \daox\ is not very sensitive to these errors, and this
X-ray-normal sample typically has smaller errors in $f_{\rm 2 keV}$
because of better photon statistics.

To investigate if this subset, the \xray\ bright \balqs, also shows
evidence for intrinsic absorption, we performed a joint-spectral fit
on the four quasars with \mbox{\daox\,$>-0.2$} to determine their average
\xray\ spectral properties.  The four \xray\ bright \balqs,
B2211$-$1915, B0029$+$0017, B1235$+$0857, and B1205$+$1436 (in order
of increasing \daox), have a total of 265 0.5--8.0~keV counts.  The
spectra were extracted using 2$\farcs$5-radius source cells from the
level~2 events files, and the arfs and rmfs were generated using the
CIAO script {\it psextract}.  Our spectral model was a simple absorbed
power-law continuum with both Galactic and intrinsic absorption by
neutral gas with solar abundances; the limited photon statistics do
not warrant a more complex model.  For joint-spectral fitting, the
parameters of interest, intrinsic \nh\ and $\Gamma$, are fit together
while the Galactic \nh\ and $z$ are fixed to the appropriate values for
each quasar.  For each quasar, the individual power-law normalizations
are free to vary (see \S4 of \citealt{gall+05} for further details).
In this low-count regime, the spectra are not binned and the fit is
performed with {\em XSPEC} by minimizing the $C$-statistic
\citep{cstat_ref}.

From the joint-spectral fitting, the best-fit model includes a
significant detection of intrinsic absorption with
\nh$=(1.95^{+0.55}_{-1.66})\times10^{22}$\cmsq\ and a normal
radio-quiet quasar \xray\ continuum with $\Gamma=2.14^{+0.55}_{-0.59}$
(the stated uncertainties are 90\% confidence for two parameters of
interest).  Excluding the radio-loud \balq\ B2211$-$1915 from the
sample increases the upper error on \nh\ from $28\%$ to $82\%$, but
otherwise does not significantly affect the best-fitting values of
$\Gamma$ and \nh.  At these redshifts, this amount of intrinsic
absorption does not substantially alter \daox, and the decrease in
\GHR\ is consistent with the fitted column density and underlying
$\Gamma$.  The additional energy resolution from spectral analysis
confirms the amount of absorption implied by the \GHR\ values
(assuming a normal radio-quiet $\Gamma\sim2$).

\begin{figure*}[t]
\plottwo{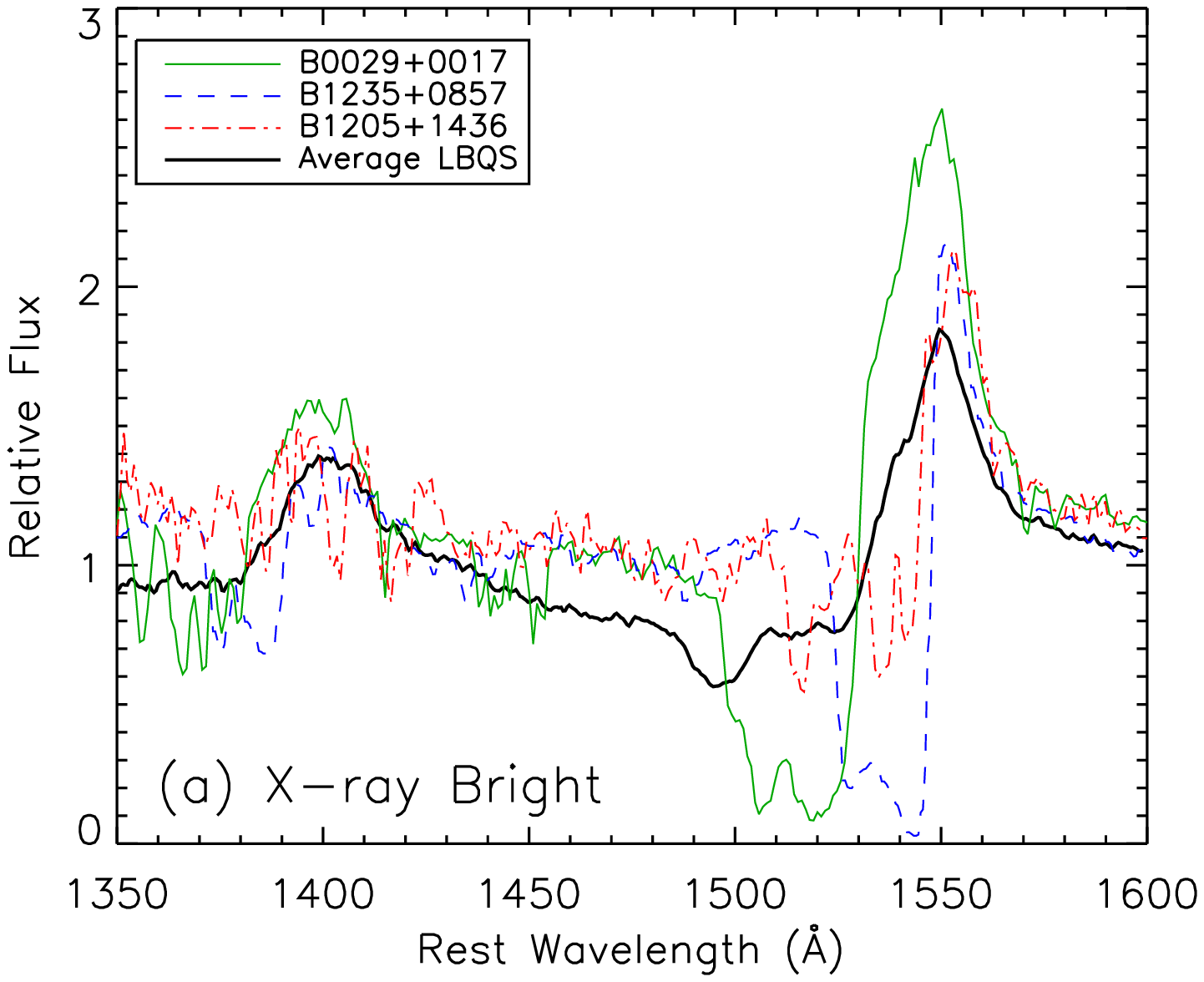}{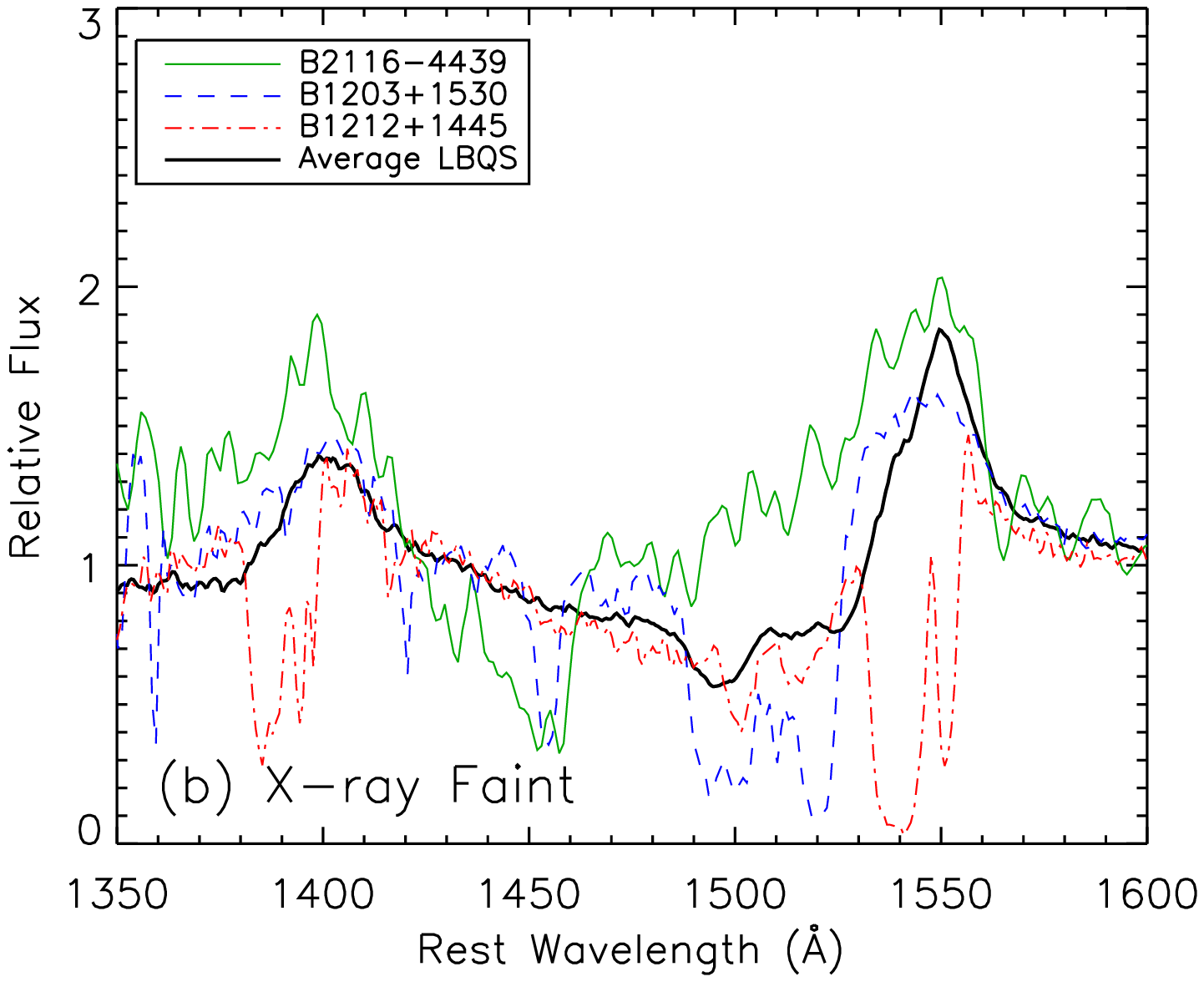}
\caption[]{ The \CIV\ regions of {\bf (a)} the three \xray\
brightest \balqs\ and {\bf (b)} the three \xray\ faintest Hi\balqs\
(\daox\,$<-0.64$) with at least moderate-quality spectra.  In both
panels, the plotted
spectra are listed in the legend in order of increasing \daox, and the average
spectrum (thick, black curve) is overplotted. Each spectrum has been
normalized to the mean flux in the 1600--1800\,\AA\ region.   
Individual and average spectra are from
\korista\ except for B2116$-$4439 \citep{lbqs_ref}. The
apparent shallow absorption structure of the average spectrum results from a smoothing
of the complex and diverse BAL troughs seen in individual objects.
All three of the X-ray brightest \balqs\ have relatively narrow,
low-velocity troughs compared to the average spectrum; their \fdeep\
values are unremarkable (see Fig.~\ref{fig:abs}d). 
\label{fig:brightspec}
}
\end{figure*}

While $\sim2\times10^{22}$\cmsq\ is on the low end of the distribution
of \nh\ for \balqs\ in general, at least two other \balqs,
PG~2112$+$059 and CSO~755, have spectroscopically measured absorption
of approximately this magnitude \citep{GaEtal2001a,ohad+05}.  The
three \xray\ brightest \balqs\ in the sample, B0029$+$0017,
B1235$+$0857, and B1205$+$1436, all have spectra from \korista, and we
plot their \CIV\ profiles as well as an average spectrum of the 27
objects in our sample observed by \korista\ in
Figure~\ref{fig:brightspec}a.  The \xray\ brightest \balqs\ have
narrow absorption troughs with low onset and maximum velocities
compared to the mean.  This is quantitatively shown by the relatively
low values of both DI and \vmax\ of these objects (see
Fig.~\ref{fig:abs}b and c), though their BIs are in the middle of the
\balq\ distribution (5263, 815, and 788\kms).  The fourth \xray\
bright (and radio-loud) \balq, B2211$-$1915, also has a relatively
narrow BAL \citep[see Fig. 1 of][]{MorrisEtal1991} with a low (for a
\balq) maximum velocity of 11,544\kms; its BI value of 27\kms\ is the
lowest in the sample.  For comparison, we also show the \CIV\ spectra
of the three X-ray faintest Hi\balqs\ with at least moderate quality
spectra (Fig.~\ref{fig:brightspec}b).  Qualitatively, their BALs are
much broader and extend to higher velocities while their BI values
(2594, 1517, and 3618\kms) are moderate.

\subsection{Comparison to the $z<0.5$ BQS Quasars}
\label{sec:bqs}

\begin{figure*}[t]
\plottwo{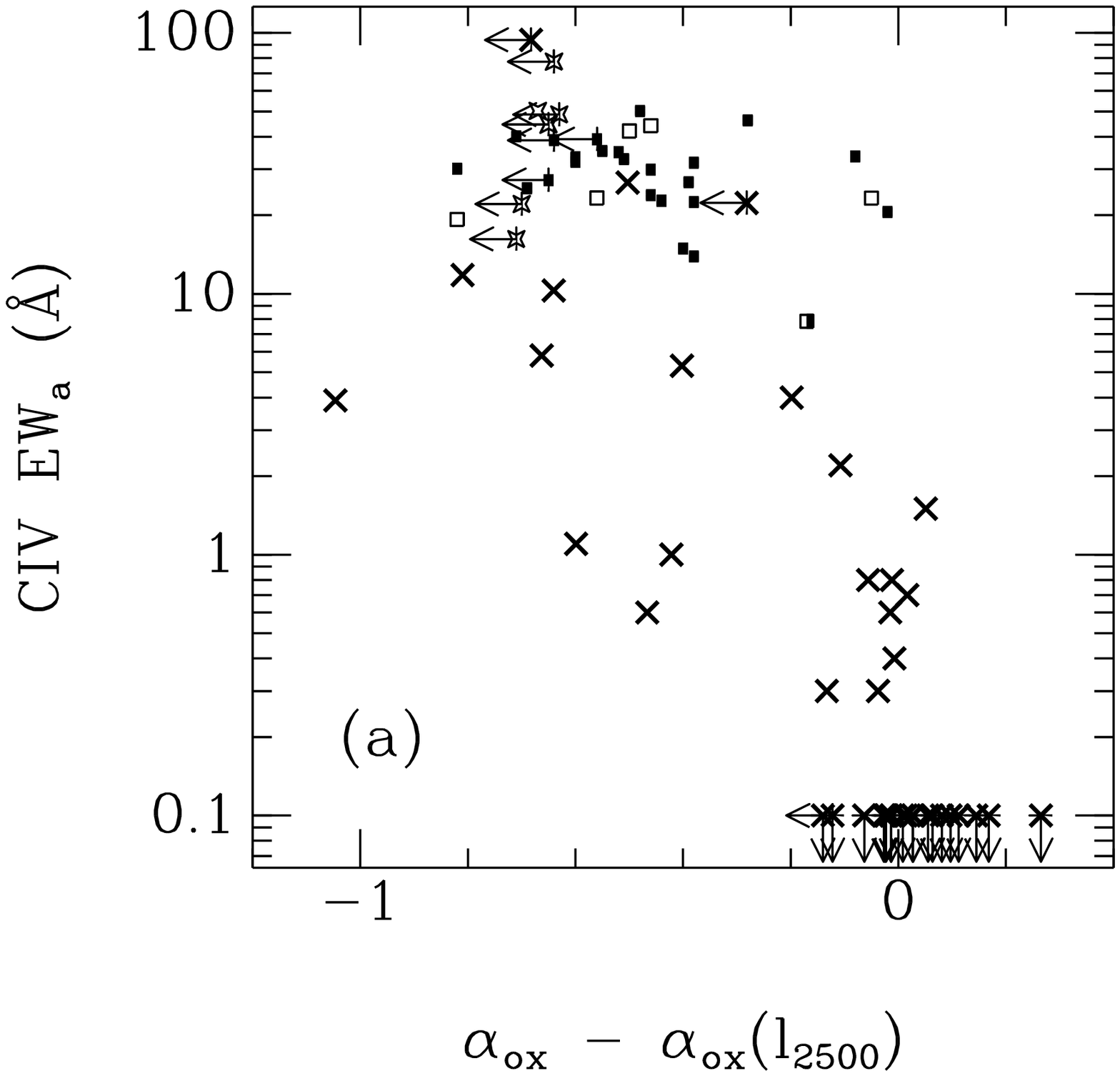}{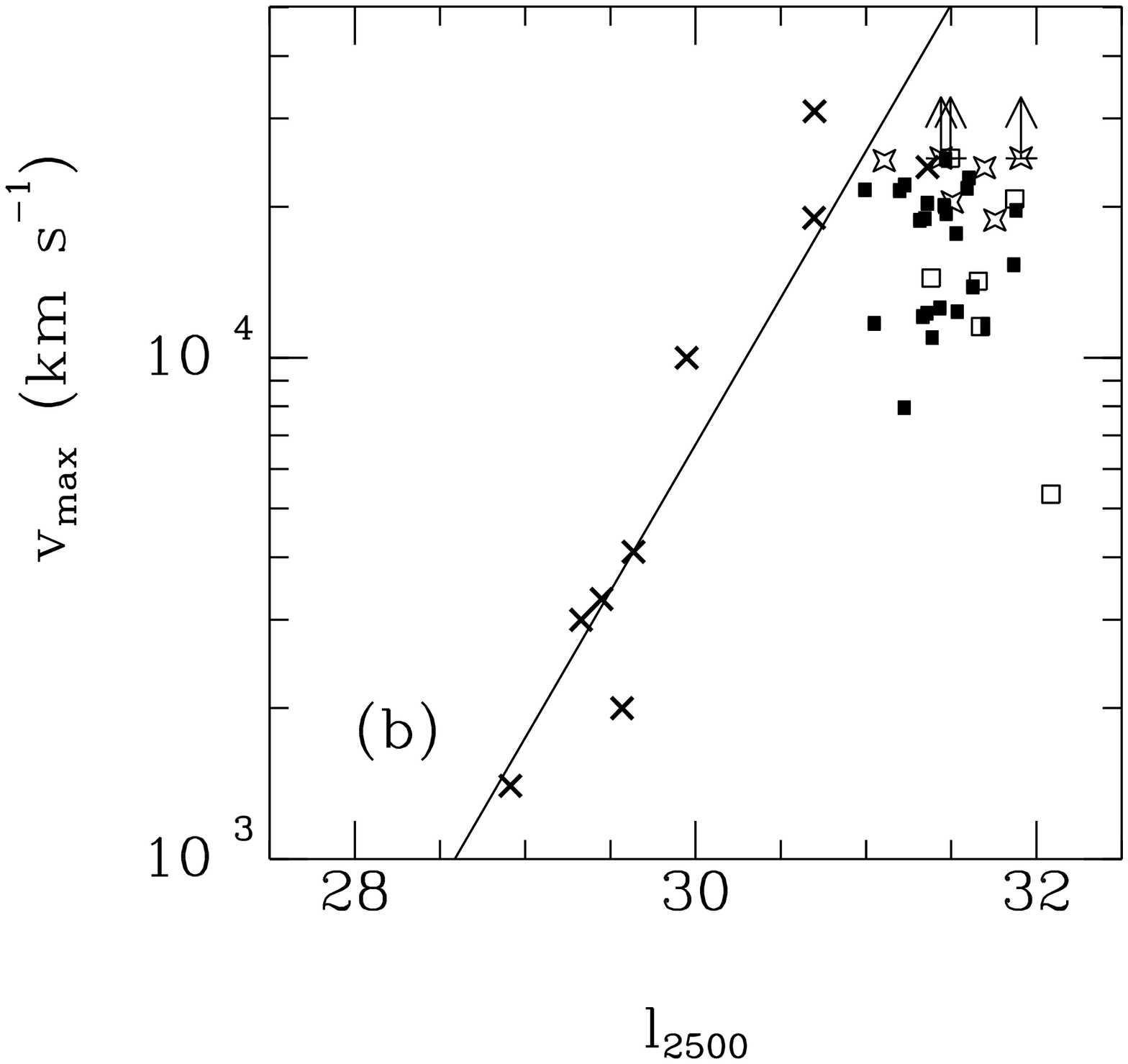}
\caption[]{Comparison of this \balq\ sample (with symbols as in
  Fig.~\ref{fig:hr}) to the $z<0.5$ BQS sample presented in
  \citet{BrLaWi2000} and \citet{LaoBra2002}.  {\bf(a)} \CIV~EW$_{\rm
  a}$ (\AA) vs. \daox; BQS sample (marked with $\times$'s) from
  \citet{BrLaWi2000}.   {\bf (b)} The parameter \vmax\ vs. $l_{2500}$.
  The solid line marks the best linear fit to the nine radio-quiet
  X-ray weak BQS quasars (marked with $\times$'s) presented in
  \citet{LaoBra2002}.
\label{fig:pg}
}
\end{figure*}

With this sample of \balqs, we can add significantly to the sample of
quasars with \xray\ weakness data and strong \uv\ absorption as
originally presented in BLW for the $z<0.5$ BQS objects.  The BLW
sample only contained a few \balqs\ but encompassed a large dynamic
range of \CIV~EW$_{\rm a}$; the majority of the BQS sample exhibited
neither weakness in \hbox{X-rays} nor \CIV\ absorption.  For this
comparison, we use \daox\ rather than \aox\ because of the large and
disparate luminosity ranges spanned by the BQS and LBQS \balq\
samples; LBQS quasars are typically \hbox{10--100} times more
luminous.  For the record, the bivariate correlation statistics of
\CIV~EW$_{\rm a}$ vs. \daox\ for the BQS sample are presented in
Table~\ref{tab:stats}.  As seen in Figure~\ref{fig:pg}a which shows
\CIV~\ew\ vs. \daox\ for both our sample and the BLW sample, the
\balqs\ occupy the region of the most extreme \CIV~EW$_{\rm a}$ (as
expected), though they do not extend to much larger values of \daox.
Though this is partially due to the detection limit of the present
survey, the mean detection of the Lo\balqs\ suggests that the \daox\
distribution would end at \daox$\sim-0.9$, probably indicating the
Compton-thick limit; some BQS objects are similarly \xray\ weak even
with notably smaller \CIV~EW$_{\rm a}$ values.

This \balq\ sample is probing the most extreme end of the \CIV\
absorption distribution, where a straightforward relationship between
weakness in soft \xray{s} and ultraviolet continuum absorption
apparently does not hold.  Given the well-documented complexity of BAL
profiles, which can include contributions from scattered light and
exhibit severe saturation \citep[e.g.,][]{OgCoMiTr1999,AravEtal2001},
the observed lack of correlation is perhaps understandable.  However,
it does seem clear that, particularly in the X-ray weakest \balqs, the
\uv\ continuum is unlikely to be obscured by the large amount of gas
blocking the X-ray continuum.

\citet{LaoBra2002} found that for a large luminosity range, \vmax\ was
significantly correlated with $M_V$ for the ten \xray\ weak objects in
their sample.  This is consistent with similar types of radiatively
driven outflows being found on all luminosity scales.  We show \vmax\
vs. $l_{\rm 2500}$ in Figure~\ref{fig:pg}b for both the \xray\ weak
quasars from \citet{LaoBra2002} as well as the LBQS \xray\ \balq\
sample.  We use $l_{\rm 2500}$ rather than $M_V$ because the former
quantity is readily available for both samples. Also, $l_{\rm 2500}$
provides a more direct and easily comparable measure of the \uv\
continuum.  The LBQS \balqs, at the high end of the luminosity
distribution, do not lie on the correlation of \citet{LaoBra2002},
though that correlation may define the maximum \vmax\ possible for a
given \uv\ luminosity.  Alternatively, the correlation may saturate at
high luminosities.  For the LBQS sample alone, \vmax\ and $l_{\rm
2500}$ are not correlated (see Table~\ref{tab:stats}); however, the
narrow range of $l_{2500}$ means this is not an ideal sample for
testing this correlation.

Though some of the LBQS sample have values of \daox\ that would not
qualify as formally \xray\ weak according to the criterion of
\citet{LaoBra2002}, this is largely an artifact of the different
bandpasses (\hbox{0.5--8.0\,keV} for \chandra\
vs. \hbox{0.1--2.4\,keV} for \rosat) and redshift ranges defining the
two samples.  The spectroscopic evidence indicates that the X-ray
bright LBQS \balqs\ are absorbed with \mbox{\nh\,$\sim10^{22}$\cmsq} (see
\S\ref{sec:joint}), similar to the amount of absorption found in some
X-ray faint BQS quasars (e.g., PG~2112$+$059).

\subsection{Implications for Disk-Wind Models}
\label{sec:diskwind}
While \xray\ spectroscopy of individual objects has in general
supported the radiatively driven disk-wind paradigm for \balqs\
(MCGV), this expansion of the population of \balqs\ with \xray\ data
offers further insight.  In the MCGV picture, equatorial \xray\
absorbing gas with \hbox{\nh$=10^{22}$--$10^{23}$\cmsq} is required
interior to the BAL gas to shield the disk wind, driven by
resonance-line pressure from ultraviolet photons, from becoming
completely ionized by the soft \xray{s} generated near the central
engine.  The range of inferred column densities for the detected
\balqs, $10^{22}$ to more than a few $10^{23}$\cmsq\ (see
Fig.~\ref{fig:hr}), is generally consistent with the shielding gas in
the MCGV models as well the hydrodynamic models of
\citet*{PrStKa2000}.  Furthermore, the mismatch of the inferred column
densities for the \xray\ absorbers and reasonable values for the \uv\
absorbers ($\lesssim10^{22}$\cmsq; e.g., \citealt{AravEtal2001})
appears to be prevalent.  This is most evident in those \balqs\ whose
\xray\ properties point toward Compton-thick absorption.  Such
obscuration clearly cannot cover a significant fraction of the \uv\
continuum source; a simple explanation for these constraints is a very
compact, thick absorber interior to the region generating the \uv\
continuum.\footnote{It is also possible that optical selection
effects (e.g., arising from the different $K$-corrections for BAL
vs. non-BAL quasars; \citealt{HewFol2003}) may play a role not yet
understood that affects the \uv/X-ray relations.}  This scenario is {\em
not} analogous to the situation for Compton-thick Seyfert~2s where the
\uv\ continuum and broad emission lines are all blocked, and a single,
cold absorber (i.e., the putative `torus') at much larger radii can
account for the observations.

Additional support for the importance of radiation pressure in driving
the \uv\ BAL wind comes from the likely connection between \vmax\ and \daox.
Specifically, the \balqs\ with the largest values of \vmax\ are
extremely X-ray weak, and the X-ray brightest \balqs\ all have
relatively narrow, low-velocity \uv\ BALs.  These empirical results
point toward a link between the column density of X-ray
absorbing gas and the maximum velocity that can be attained by the
\uv\ BAL wind.  

Generically, the terminal velocity of outflowing material is set to
some extent by the launching radius (e.g., \citealt{ChBrGa2003} and
references therein). Gas launched from smaller radii must attain
higher velocities in order to escape from the vicinity of the black
hole; this material is also exposed to higher photon densities.  We
speculate that the thickness of the X-ray absorber (with more negative
values of \daox\ indicating larger X-ray opacities) determines the
innermost radius from which the BAL wind could potentially be
launched; thus, without a very thick shield, material at smaller radii
will be too ionized by soft X-rays to be driven effectively by \uv\
line pressure.  This situation, where the amount of shielding affects
the velocity of the outflow, is generally consistent with the models
of \citet{everett05} of accretion-disk winds driven by both
magnetocentrifugal and radiative forces (J. Everett, 2005,
priv. comm.).  Note that this scenario would not apply to the
relativistic \xray\ BALs seen in a few objects
\citep[e.g.,][]{ChBrGa2003}; such gas, inferred to be very highly
ionized, is unlikely to be accelerated by \uv-line pressure. The large
scatter in \vmax\ corresponding to the most negative \daox\ values may
indicate that a thick absorber is a necessary but not sufficient
condition for seeing high velocity outflows along the \uv\ line of
sight.  Additional observations are certainly required to investigate
this possibility.  

More subtly, the lack of correlation of \daox\ with DI (a proxy for
the minimum velocity of absorption) implies that a straightforward
transverse acceleration model for large inclination angles does not
hold.  As explained in \citet{Goodrich1997}, from pure orientation
effects, larger column densities of gas (resulting in more negative
values of \daox) would be expected along lines of sight skimming the
disk.  The wind seen through these observing angles would have a
larger transverse velocity component from local radiation pressure
initially acting primarily perpendicular to the accretion disk.  After
this initial launching stage, photons from the inner accretion disk
accelerate the gas radially.  The net result of this launching process
would be lower line-of-sight onset velocities of absorption and thus
smaller values of DI at large inclination angles.  At smaller
inclination angles where the line of sight intercepts the wind after
it has been lifted from the disk, the outflow is more radial, and
therefore DI would be larger. The lack of correlation found between DI
and \daox\ in our data does not support this simple picture with a
wind geometry that `turns over' (e.g., see Fig.~1
\altcite{MuChGrVo1995}).  The more complex hydrodynamic wind models
presented in \citet{PrStKa2000} and \citet{pk2004} do not make
specific predictions for the relationship between absorption onset
velocity and X-ray absorption column density (D. Proga, 2005,
priv. comm.), and so it is unclear if they are consistent with our
result.

\section{Conclusions and Summary}
\label{sec:conc}

The sample of \balqs\ studied in this paper offers the advantages of
being much larger and more homogeneous than those in earlier hard-band
surveys.  Specifically, our sample ranges over relatively narrow spans
of both redshift ($z=1.42-2.90$) and ultraviolet luminosity (a factor
of $\approx12$). In addition, the utilization of \uv\ spectroscopic
data for all objects has enabled more specific quantitative
comparisons between properties in different wavelength regimes than
has been possible previously.

We briefly summarize our conclusions below:

\begin{enumerate}

\item{We confirm and extend previous work finding \balqs\ to be
  generally X-ray weak.  However, from the days of \rosat\ and \asca,
  the detection fraction of \balqs\ has increased substantially. Of
  the \total\ observed LBQS \balqs, \detected\ were detected for a
  detection fraction of 77\%.  Given the sensitivity of \chandra,
  meaningful upper limits on \xray\ flux and \aox\ have been set for
  those that were not detected.}

\item{The distributions of $\Delta$\aoxc\ and \GHR\ versus \daox\ both
  support the hypothesis that \xray\ weakness in \balqs\ arises from
  intrinsic absorption with column densities of
  \hbox{$\approx$(0.1--10)$\times10^{23}$\cmsq}.  Furthermore, the
  \GHR\ versus \daox\ distribution is not consistent with simple
  neutral absorption and supports previous claims from \xray\
  spectroscopy for complex absorbers.  Specifically, for a given
  \daox, more soft photons are reaching the observer than would be
  expected with a simple neutral absorber, and partially covering
  and/or ionized absorbers can explain this.  Even the \balqs\ with
  normal \daox\ values in this sample show evidence from
  joint-spectral fitting for more moderate amounts of absorption,
  \mbox{\nh\,$\sim10^{22}$\cmsq}.  The non-detected objects are
  candidates for Compton-thick \xray\ obscuration.}

\item{Three measures of the \uv-absorber properties, BI, DI, and
  \fdeep, are uncorrelated with the \xray\ weakness in this sample.
  This lack of a straightforward relationship between the \uv\ and
  \xray\ absorption as characterized by these parameters is difficult
  to account for with a single absorbing structure that obscures the
  \uv\ and \xray\ continuum in the same manner.  Furthermore, the
  lack of correlation of DI and \daox\ is inconsistent with
  predictions from disk-wind models with a significant transverse
  velocity component at large inclination angles.  A fourth property,
  \vmax, is correlated with \daox\ at the $\sim3\sigma$ level.
  Though this correlation requires confirmation with larger sample
  sizes, it is nonetheless evident that the \balqs\ with the largest
  outflow velocities are typically \xray\ weakest.  This plausibly
  implies that very thick \xray\ absorption is required to achieve the
  highest outflow velocities seen in the \uv, a natural consequence of
  a radiatively driven disk-wind model that requires shielding to
  prevent overionization of the wind.}

\item{The large fraction of non-\xray\ detected Lo\balqs, $\sim$80\%,
  and their average value of \daox$\sim$--0.9 estimated from stacking
  analysis, point toward extreme \xray\ absorption in this subset of
  \balqs.  The Lo\balq\ Mrk~231 is already known to show spectroscopic
  evidence for Compton-thick \xray\ absorption, and this appears to be
  typical of the class.  In fact, it may be the case that (near)
  Compton-thick absorption is a prerequisite for seeing \MgII\ BALs
  along the line of sight to the observer. Given these extreme \xray\
  and ultraviolet absorption properties, it is important to
  differentiate the samples of low and high-ionization \balqs\ when
  trying to generalize about the population of \balqs\ as a whole.}

\end{enumerate}

In the near future, the \xray\ brightest BAL quasars in this sample
are viable candidates for spectroscopy with \xmm\ for further
investigation into the nature of their \xray\ absorption.  In
particular, those with \chandra\ count rates
$\gtrsim1\times10^{-3}$\,ct\,\persec\ will provide
\hbox{$\sim2000$--3000} EPIC counts in 40--60\,ks observations.
However, the three objects meeting this criterion are a small fraction
of the total sample, and so systematic follow-up will likely require
the next generation of large effective area \xray\ observatories.

Though it is probably not the case for the majority of \balqs, in the
\xray\ brightest \balqs\ the \uv\ and \xray\ absorbers may be
identical.  These objects will be excellent candidates for high
resolution \xray\ spectroscopy with the next generation of \xray\
observatories to compare the ionization and velocity structure of the
\uv\ and \xray\ absorbing gas.  However, it is clear that they are not
representative of the class as a whole.  For the more typical objects,
specific models of the profiles of \xray\ absorption lines compared to
\uv\ absorption-line profiles will provide important predictions
for future high-throughput spectroscopic missions.

\acknowledgements 

We thank the following individuals for generously providing electronic
data: Fred Hamann (\CIV~EW$_{\rm a}$ measurements), Paul Hewett (LBQS
spectra), Kirk Korista and Simon Morris (spectra from \korista), Aaron
Steffen (BQS quasar data), and Iskra Strateva (SDSS \aox\ data). Pat Hall
and our anonymous referee provided constructive comments that improved
this paper. We also acknowledge the Sloan Digital Sky Survey
(http://www.sdss.org).  This work was made possible by \chandra\
\xray\ Center grants GO1-2105X, GO4-5113X, and NASA grant NAS8-38252
(PI G.\ P. Garmire) that supports the ACIS Instrument Team.  Support
for SCG was provided by NASA through the {\em Spitzer} Fellowship
Program, under award 1256317.  WNB acknowledges NASA LTSA grant
NAG5--13035.



\begin{thebibliography}{70}
\expandafter\ifx\csname natexlab\endcsname\relax\def\natexlab#1{#1}\fi
\expandafter\ifx\csname url\endcsname\relax
  \def\url#1{{\tt #1}}\fi
\expandafter\ifx\csname urlprefix\endcsname\relax\def\urlprefix{URL }\fi
\providecommand{\eprint}[2][]{\url{#2}}

\bibitem[\protect\astroncite{{Aldcroft} \& {Green}}{2003}]{AlGr2003}
{Aldcroft}, T.~L. \& {Green}, P.~J. 2003, \apj, 592, 710

\bibitem[\protect\astroncite{{Arav} et~al.}{2001}]{AravEtal2001}
{Arav}, N., et~al. 2001, \apj, 561, 118

\bibitem[\protect\astroncite{{Arnaud}}{1996}]{xspec_ref}
{Arnaud}, K.~A. 1996, in ASP Conf. Ser. 101: Astronomical Data Analysis
  Software and Systems V, eds. G.~Jacoby \& J.~Barnes, vol.~5, 17

\bibitem[\protect\astroncite{{Avni} \& {Tananbaum}}{1986}]{AvTa1986}
{Avni}, Y. \& {Tananbaum}, H. 1986, \apj, 305, 83

\bibitem[\protect\astroncite{{Bauer} et~al.}{2004}]{BauerEtal2004}
{Bauer}, F.~E., {Alexander}, D.~M., {Brandt}, W.~N., {Schneider}, D.~P.,
  {Treister}, E., {Hornschemeier}, A.~E., \& {Garmire}, G.~P. 2004, \aj, 128,
  2048

\bibitem[\protect\astroncite{{Braito} et~al.}{2004}]{braito+04}
{Braito}, V., et~al. 2004, \aap, 420, 79

\bibitem[\protect\astroncite{{Brandt} et~al.}{2000}]{BrLaWi2000}
{Brandt}, W.~N., {Laor}, A., \& {Wills}, B.~J. 2000, \apj, 528, 637 (BLW)z

\bibitem[\protect\astroncite{{Brandt} et~al.}{2001}]{stacking_ref}
{Brandt}, W.~N., et~al. 2001, \aj, 122, 1

\bibitem[\protect\astroncite{{Cash}}{1979}]{cstat_ref}
{Cash}, W. 1979, \apj, 228, 939

\bibitem[\protect\astroncite{{Chartas} et~al.}{2003}]{ChBrGa2003}
{Chartas}, G., {Brandt}, W.~N., \& {Gallagher}, S.~C. 2003, \apj, 595, 85

\bibitem[\protect\astroncite{{Chartas} et~al.}{2002}]{ChBrGaGa2002}
{Chartas}, G., {Brandt}, W.~N., {Gallagher}, S.~C., \& {Garmire}, G.~P. 2002,
  \apj, 579, 169

\bibitem[\protect\astroncite{{Condon} et~al.}{1998}]{nvss_ref}
{Condon}, J.~J., {Cotton}, W.~D., {Greisen}, E.~W., {Yin}, Q.~F., {Perley},
  R.~A., {Taylor}, G.~B., \& {Broderick}, J.~J. 1998, \aj, 115, 1693

\bibitem[\protect\astroncite{{Cristiani} \& {Vio}}{1990}]{CrVi1990}
{Cristiani}, S. \& {Vio}, R. 1990, \aap, 227, 385

\bibitem[\protect\astroncite{{Dickey} \& {Lockman}}{1990}]{HI_ref}
{Dickey}, J.~M. \& {Lockman}, F.~J. 1990, \araa, 28, 215

\bibitem[\protect\astroncite{Everett}{2005}]{everett05} 
Everett, J.~E.\ 2005, \apj, 631, 689 

\bibitem[\protect\astroncite{{Feigelson} \& {Nelson}}{1985}]{FeNe1985}
{Feigelson}, E.~D. \& {Nelson}, P.~I. 1985, \apj, 293, 192

\bibitem[\protect\astroncite{{Fitzpatrick} \& {Massa}}{1999}]{fm99} 
Fitzpatrick, E.~L., \& Massa, D.\ 1999, \apj, 525, 1011 
 
\bibitem[\protect\astroncite{{Freeman} et~al.}{2002}]{wavdet_ref}
{Freeman}, P.~E., {Kashyap}, V., {Rosner}, R., \& {Lamb}, D.~Q. 2002, \apjs,
  138, 185

\bibitem[\protect\astroncite{{Gallagher} et~al.}{2002}]{GaBrChGa2002}
{Gallagher}, S.~C., {Brandt}, W.~N., {Chartas}, G., \& {Garmire}, G.~P. 2002,
  \apj, 567, 37

\bibitem[\protect\astroncite{{Gallagher} et~al.}{2001}]{GaEtal2001a}
{Gallagher}, S.~C., {Brandt}, W.~N., {Laor}, A., {Elvis}, M., {Mathur}, S.,
  {Wills}, B.~J., \& {Iyomoto}, N. 2001, \apj, 546, 795

\bibitem[\protect\astroncite{{Gallagher} et~al.}{1999}]{GaEtal1999}
{Gallagher}, S.~C., {Brandt}, W.~N., {Sambruna}, R.~M., {Mathur}, S., \&
  {Yamasaki}, N. 1999, \apj, 519, 549

\bibitem[\protect\astroncite{{Gallagher} et~al.}{2004}]{gall+04}
{Gallagher}, S.~C., {Brandt}, W.~N., {Wills}, B.~J., {Charlton}, J.~C.,
  {Chartas}, G., \& {Laor}, A. 2004, \apj, 603, 425

\bibitem[\protect\astroncite{{Gallagher} et~al.}{2005}]{gall+05}
{Gallagher}, S.~C., {Richards}, G.~T., {Hall}, P.~B., {Brandt}, W.~N.,
  {Schneider}, D.~P., \& {Vanden Berk}, D.~E. 2005, \aj, 129, 567

\bibitem[\protect\astroncite{{Garmire} et~al.}{2003}]{acis_ref}
{Garmire}, G.~P., {Bautz}, M.~W., {Ford}, P.~G., {Nousek}, J.~A., \& {Ricker},
  G.~R. 2003, in Proc. SPIE Vol. 4851: X-Ray and Gamma-Ray Telescopes and
  Instruments for Astronomy, eds. J.~E. Tr\"umper \& H.~D. Tananbaum, 28--44

\bibitem[\protect\astroncite{{Gehrels}}{1986}]{Gehrels}
{Gehrels}, N. 1986, \apj, 303, 336

\bibitem[\protect\astroncite{{George} et~al.}{2000}]{GeoEtal2000}
{George}, I.~M., {Turner}, T.~J., {Yaqoob}, T., {Netzer}, H., {Laor}, A.,
  {Mushotzky}, R.~F., {Nandra}, K., \& {Takahashi}, T. 2000, \apj, 531, 52

\bibitem[\protect\astroncite{{Goodrich}}{1997}]{Goodrich1997}
{Goodrich}, R.~W. 1997, \apj, 474, 606

\bibitem[\protect\astroncite{{Green} et~al.}{2001}]{GreenEtal2001}
{Green}, P.~J., {Aldcroft}, T.~L., {Mathur}, S., {Wilkes}, B.~J., \& {Elvis},
  M. 2001, \apj, 558, 109

\bibitem[\protect\astroncite{{Green} \& {Mathur}}{1996}]{GrMa1996}
{Green}, P.~J. \& {Mathur}, S. 1996, \apj, 462, 637

\bibitem[\protect\astroncite{Green et~al.}{1995}]{GrEtal1995}
Green, P.~J., et~al. 1995, \apj, 450, 51

\bibitem[\protect\astroncite{{Grupe} et~al.}{2003}]{GrMaEl2003}
{Grupe}, D., {Mathur}, S., \& {Elvis}, M. 2003, \aj, 126, 1159

\bibitem[\protect\astroncite{{Hamann} et~al.}{1993}]{HaKoMo1993}
{Hamann}, F., {Korista}, K.~T., \& {Morris}, S.~L. 1993, \apj, 415, 541

\bibitem[\protect\astroncite{{Hewett} \& {Foltz}}{2003}]{HewFol2003}
{Hewett}, P.~C. \& {Foltz}, C.~B. 2003, \aj, 125, 1784

\bibitem[\protect\astroncite{{Hewett} et~al.}{1995}]{lbqs_ref}
{Hewett}, P.~C., {Foltz}, C.~B., \& {Chaffee}, F.~H. 1995, \aj, 109, 1498

\bibitem[\protect\astroncite{{Hewett} et~al.}{2001}]{HeFoCh2001}
--- 2001, \aj, 122, 518

\bibitem[\protect\astroncite{{Isobe} et~al.}{1986}]{IsFeNe1986}
{Isobe}, T., {Feigelson}, E.~D., \& {Nelson}, P.~I. 1986, \apj, 306, 490

\bibitem[\protect\astroncite{{Ivezi{\'c}} et~al.}{2002}]{ivezic2002}
{Ivezi{\'c}}, {\v Z}., et~al. 2002, \aj, 124, 2364

\bibitem[\protect\astroncite{{K\"onigl} \& {Kartje}}{1994}]{KoKa1994}
{K\"onigl}, A. \& {Kartje}, J.~F. 1994, \apj, 434, 446

\bibitem[\protect\astroncite{{Kopko} et~al.}{1994}]{KoTuEs1994}
{Kopko}, M., {Turnshek}, D.~A., \& {Espey}, B.~R. 1994, in IAU Symp. 159:
  {Multi}-Wavelength Continuum Emission of AGN, eds. T.~Courvoisier \&
  A.~Blecha, vol. 159 (Dordrecht: Kluwer), 450

\bibitem[\protect\astroncite{{Korista} et~al.}{1993}]{KoVoMoWe1993}
{Korista}, K.~T., {Voit}, G.~M., {Morris}, S.~L., \& {Weymann}, R.~J. 1993,
  \apjs, 88, 357, \eprint{(KVMW)}

\bibitem[\protect\astroncite{{Kraft} et~al.}{1991}]{kbn}
{Kraft}, R.~P., {Burrows}, D.~N., \& {Nousek}, J.~A. 1991, \apj, 374, 344

\bibitem[\protect\astroncite{{La~Valley} et~al.}{1992}]{LaIsFe1992}
{La~Valley}, M., {Isobe}, T., \& {Feigelson}, E. 1992, in ASP Conf. Ser. 25:
  Astronomical Data Analysis Software and Systems I, eds. D.~M. Worrall,
  C.~Biemesderfer, \& J.~Barnes, vol.~1, 245

\bibitem[\protect\astroncite{{Lamy} \& {Hutsem{\' e}kers}}{2004}]{LaHu2004}
{Lamy}, H. \& {Hutsem{\' e}kers}, D. 2004, \aap, 427, 107

\bibitem[\protect\astroncite{{Laor} \& {Brandt}}{2002}]{LaoBra2002}
{Laor}, A. \& {Brandt}, W.~N. 2002, \apj, 569, 641

\bibitem[\protect\astroncite{Lyons}{1991}]{Lyons1991}
Lyons, L. 1991, \emph{Data Analysis for Physical Science Students} (Cambridge:
  Cambridge University Press)

\bibitem[\protect\astroncite{{Marshall} et~al.}{2004}]{acis_contam}
{Marshall}, H.~L., {Tennant}, A., {Grant}, C.~E., {Hitchcock}, A.~P., {O'Dell},
  S.~L., \& {Plucinsky}, P.~P. 2004, in X-Ray and Gamma-Ray Instrumentation for
  Astronomy XIII., Proceedings of the SPIE, Volume 5165, 497--508

\bibitem[\protect\astroncite{{Mathur} et~al.}{2000}]{MathurEtal2000}
{Mathur}, S., et~al. 2000, \apjl, 533, L79

\bibitem[\protect\astroncite{{Morris} et~al.}{1991}]{MorrisEtal1991}
{Morris}, S.~L., {Weymann}, R.~J., {Anderson}, S.~F., {Hewett}, P.~C.,
  {Francis}, P.~J., {Foltz}, C.~B., {Chaffee}, F.~H., \& {MacAlpine}, G.~M.
  1991, \aj, 102, 1627

\bibitem[\protect\astroncite{{Murray} et~al.}{1995}]{MuChGrVo1995}
{Murray}, N., {Chiang}, J., {Grossman}, S.~A., \& {Voit}, G.~M. 1995, \apj,
  451, 498, \eprint{(MCGV)}

\bibitem[\protect\astroncite{{Ogle} et~al.}{1999}]{OgCoMiTr1999}
{Ogle}, P.~M., {Cohen}, M.~H., {Miller}, J.~S., {Tran}, H.~D., {Goodrich},
  R.~W., \& {Martel}, A.~R. 1999, \apjs, 125, 1

\bibitem[\protect\astroncite{{Page} et~al.}{2005}]{PageEtal2005}
{Page}, K.~L., {Reeves}, J.~N., {O'Brien}, P.~T., \& {Turner}, M.~J.~L. 2005,
  \mnras, 898

\bibitem[\protect\astroncite{Press et~al.}{1997}]{Recipes}
Press, W.~H., Teukolsky, S.~A., Vetterling, W.~T., \& Flannery, B.~P. 1997,
  \emph{Numerical Recipes in C: The Art of Scientific Computing}, 2nd ed.
  (Cambridge: Cambridge University Press)

\bibitem[\protect\astroncite{{Proga} \& {Kallman}}{2004}]{pk2004}
{Proga}, D. \& {Kallman}, T.~R. 2004, \apj, 616, 688

\bibitem[\protect\astroncite{{Proga} et~al.}{2000}]{PrStKa2000}
{Proga}, D., {Stone}, J.~M., \& {Kallman}, T.~R. 2000, \apj, 543, 686

\bibitem[\protect\astroncite{{Reeves} \& {Turner}}{2000}]{ReTu2000}
{Reeves}, J.~N. \& {Turner}, M. J.~L. 2000, \mnras, 316, 234

\bibitem[\protect\astroncite{{Reichard} et~al.}{2003}]{reichar+03}
{Reichard}, T.~A., et~al. 2003, \aj, 126, 2594

\bibitem[\protect\astroncite{{Richards} et~al.}{2002}]{ric+02}
{Richards}, G.~T., {Vanden Berk}, D.~E., {Reichard}, T.~A., {Hall}, P.~B.,
  {Schneider}, D.~P., {SubbaRao}, M., {Thakar}, A.~R., \& {York}, D.~G. 2002,
  \aj, 124, 1

\bibitem[\protect\astroncite{{Sabra} \& {Hamann}}{2001}]{SaHa2001}
{Sabra}, B.~M. \& {Hamann}, F. 2001, \apj, 563, 555

\bibitem[\protect\astroncite{{Schmidt} \& {Green}}{1983}]{pg_ref}
{Schmidt}, M. \& {Green}, R.~F. 1983, \apj, 269, 352

\bibitem[\protect\astroncite{Shemmer et~al.}{2005}]{ohad+05}
Shemmer, O., Brandt, W., Gallagher, S., Vignali, C., Boller, T., Chartas, G.,
  \& Comastri, A. 2005, \aj, in press, \eprint{astro-ph/0509146}

\bibitem[\protect\astroncite{{Sirola} et~al.}{1998}]{SirolaEtal1998}
{Sirola}, C.~J., et~al. 1998, \apj, 495, 659

\bibitem[\protect\astroncite{{Sprayberry} \& {Foltz}}{1992}]{SpFo1992}
{Sprayberry}, D. \& {Foltz}, C.~B. 1992, \apj, 390, 39

\bibitem[\protect\astroncite{{Stocke} et~al.}{1992}]{StMoWeFo1992}
{Stocke}, J.~T., {Morris}, S.~L., {Weymann}, R.~J., \& {Foltz}, C.~B. 1992,
  \apj, 396, 487

\bibitem[\protect\astroncite{{Strateva} et~al.}{2005}]{Strateva+2005}
{Strateva}, I.~V., {Brandt}, W.~N., {Schneider}, D.~P., {Vanden Berk}, D.~G.,
  \& {Vignali}, C. 2005, \aj, 130, 387

\bibitem[\protect\astroncite{{Tananbaum} et~al.}{1979}]{aox_ref}
{Tananbaum}, H., et~al. 1979, \apjl, 234, L9

\bibitem[\protect\astroncite{{Vanden Berk} et~al.}{2001}]{VandenBerk2001}
{Vanden Berk}, D.~E., et~al. 2001, \aj, 122, 549

\bibitem[\protect\astroncite{{Vignali} et~al.}{2003}]{VignaliEtal2003}
{Vignali}, C., {Brandt}, W.~N., \& {Schneider}, D.~P. 2003, \aj, 125, 433

\bibitem[\protect\astroncite{{Weisskopf} et~al.}{2002}]{chandra_ref}
{Weisskopf}, M.~C., {Brinkman}, B., {Canizares}, C., {Garmire}, G., {Murray},
  S., \& {Van Speybroeck}, L.~P. 2002, \pasp, 114, 1

\bibitem[\protect\astroncite{{Weymann} et~al.}{1991}]{WeMoFoHe1991}
{Weymann}, R.~J., {Morris}, S.~L., {Foltz}, C.~B., \& {Hewett}, P.~C. 1991,
  \apj, 373, 23, \eprint{(WMHF)}

\bibitem[\protect\astroncite{{White} et~al.}{1997}]{first_ref}
{White}, R.~L., {Becker}, R.~H., {Helfand}, D.~J., \& {Gregg}, M.~D. 1997,
  \apj, 475, 479

\bibitem[\protect\astroncite{{Wilkes} et~al.}{1994}]{WilkesEtal1994}
{Wilkes}, B.~J., {Tananbaum}, H., {Worrall}, D.~M., {Avni}, Y., {Oey}, M.~S.,
  \& {Flanagan}, J. 1994, \apjs, 92, 53

\bibitem[\protect\astroncite{{York} et~al.}{2000}]{sdss_ref}
{York}, D.~G., et~al. 2000, \aj, 120, 1579

\end{thebibliography}

\clearpage
\begin{landscape}
\clearpage
 \begin{deluxetable}{lcccrcccrrccr}
\tabletypesize{\scriptsize}
\tablewidth{0pt}
\tablecaption{Observed Targets\label{tab:opt}}
\tablehead{
\colhead{Name} &
\colhead{$z$\tablenotemark{a}}&
\colhead{\bj\tablenotemark{a}} &
\colhead{$f_{2500}$\tablenotemark{b}} &
\colhead{\auv\tablenotemark{c}} &
\colhead{\nh\tablenotemark{d}} &
\colhead{BI\tablenotemark{e}} &
\colhead{\CIV\ EW$_{\rm a}$} &
\colhead{\vmax\tablenotemark{f}} &
\colhead{\fdeep\tablenotemark{g}} &
\colhead{BAL} &
\colhead{$R^{\star}/R_{i}$\tablenotemark{i}} &
\colhead{Notes\tablenotemark{j}} \\
\colhead{(LBQS B)} &
\colhead{} &
\colhead{} &
\colhead{($10^{-17} f_\lambda$)} &
\colhead{} &
\colhead{($10^{20}$ \cmsq)} &
\colhead{(\kms)} &
\colhead{(\AA)} &
\colhead{(\kms)} &
\colhead{} &
\colhead{Type\tablenotemark{h}} &
\colhead{} &
\colhead{} 
}
\startdata
0004$+$0147     & 1.710 & 18.13 &  24.7 & $ -1.06$ &  3.01 &    255 & 16.2  &$>25000$ &  0.13 &  Lo & $<-0.06$/$\cdots$   &              K,S,LH \\
0019$+$0107     & 2.130 & 18.09 &  19.2 & $ -1.07$ &  3.22 &   2305 & 23.9 &  13849 &   0.44 & Hi &  $<-0.16$/$\cdots$   &         K,W,S,LH,H \\
0021$-$0213     & 2.293 & 18.68 &  10.3 & $ -1.64$ &  2.95 &   5179 & 40.2  &  20138 &   0.51 & Hi &  $< 0.11$/$\cdots$   &         K,W,S,LH,H \\
0025$-$0151     & 2.076 & 18.06 &  19.3 & $ -1.34$ &  2.85 &   2878 & 32.0  &  21765 &   0.15 & Hi &  $<-0.02$/$\cdots$   &         K,W,S,LH,H \\
0029$+$0017     & 2.253 & 18.64 &   8.7 & $ -1.72$ &  2.40 &   5263 & 33.6  &  12267 &   0.53 & Hi &  $< 0.15$/$<0.80$  	 &    K,W,SDSS,S,LH,H \\
\\								    
0051$-$0019     & 1.713 & 18.67 &  15.1 & $ -1.25$ &  3.22 &   3244 & 46.3  &  22113 &   0.53 & Hi &  $\cdots$/$<0.61$    &      K,SDSS(sp),LH \\
0054$+$0200     & 1.872 & 18.41 &  16.6 & $ -1.39$ &  3.11 &    498 & 22.7  &  10970 &   0.23 & Hi &  $\cdots$/$<0.97$\tablenotemark{i}	 &               K,LH \\
0059$-$2735     & 1.593 & 18.13 &  35.9 & $ -1.34$ &  1.99 &  11053 & 77.5  &  20427 &   0.81 & Lo &  $<-0.15$/$\cdots$   &       K,W,S,LH,H,G \\
0109$-$0128     & 1.758 & 18.32 &  17.1 & $ -1.20$ &  4.07 &    399 & 35.3  &  18794 &   0.25 & Hi &  $\cdots$/$<0.53$    &        SDSS(sp),LH \\
1029$-$0125     & 2.029 & 18.43 &  17.9 & $ -1.27$ &  4.80 &   1848 & 26.8  &  17685 &   0.43 & Hi &  $<-0.10$/$\cdots$   &         K,W,S,LH,H \\
\\								    
1133$+$0214     & 1.468 & 18.38 &  22.6 & $ -2.23$ &  2.61 &   1950 & 22.5  &  21547 &   0.00 & Hi &  $\cdots$/$<0.47$    &      SDSS(sp),LBQS \\
1203$+$1530     & 1.628 & 18.70 &  11.7 & $ -2.11$ &  2.80 &   1517 & 25.4  &  11702 &   0.04 & Hi &  $\cdots$/$<0.67$    &       SDSS,LH,LBQS \\
1205$+$1436     & 1.643 & 18.38 &  17.0 & $ -1.60$ &  2.59 &    788 & 20.6  &   7947 &   0.18 & Hi &  $\cdots$/$<0.57$    &    K,W,SDSS,S,LH,H \\
1208$+$1535     & 1.961 & 17.93 &  13.5 & $ -1.76$ &  2.67 &   4545 & 29.9  &  20325 &   0.20 & Hi &  $< 0.13$/$<0.61$    &    K,W,SDSS,S,LH,H \\
1212$+$1445     & 1.627 & 17.87 &  31.0 & $ -1.13$ &  2.69 &   3618 & 38.8  &  19368 &   0.20 & Hi &   $ 0.01$/$<0.34$    &    K,W,SDSS,S,LH,H \\
\\								   
1216$+$1103     & 1.620 & 18.28 &  22.9 & $ -2.08$ &  2.15 &   4791 & 33.3  &  12076 &   0.67 & Hi &  $< 0.04$/$<0.43$    &K,W,SDSS(sp),S,LH,H \\
1230$+$1705     & 1.420 & 18.44 &  15.7 & $ -2.63$ &  2.26 &   2945 & 34.9  &  21617 &   0.18 & Hi &   $ 0.12$/$\cdots$   &    S,LH,LBQS \\
1231$+$1320     & 2.380 & 18.84 &  26.1 & $ -0.26$ &  1.86 &   3473 & 22.1  &$>25000$ &  0.09 &  Lo & $<-0.41$/$<0.35$    &     K,W,SDSS,S,LH,H \\
1235$+$0857     & 2.898 & 18.17 &  21.0 & $ -1.47$ &  1.68 &    815 & 23.3  &   5339 &   0.69 &  ? &   $-0.12$/$\cdots$   &    K,W,SDSS,S,LH,H \\
1235$+$1453     & 2.699 & 18.56 &   5.2 & $ -1.33$ &  2.37 &   2657 & 19.3  &  14414 &   0.45 &  ? &  $< 0.42$/$<0.27$    &    K,W,SDSS,S,LH,H \\
\\								    
1239$+$0955     & 2.013 & 18.38 &  18.6 & $ -1.75$ &  1.66 &    708 & 13.9  &  12355 &   0.38 & Hi &  $\cdots$/$<0.46$    &      K,SDSS,S,LH,H \\
1240$+$1607     & 2.360 & 18.84 &   8.9 & $ -0.75$ &  2.16 &   2867 & 32.9  &  12568 &   0.32 & Hi &  $< 0.15$/$<0.81$    &    K,W,SDSS,S,LH,H \\
1243$+$0121     & 2.796 & 18.50 &  14.4 & $ -0.66$ &  1.76 &   5953 & 42.1  &  20718 &   0.47 &  ? &  $<-0.16$/$<0.58$    &K,W,SDSS(sp),S,LH,H \\
1314$+$0116     & 2.686 & 18.65 &  10.0 & $ -1.70$ &  1.98 &   2626 & 23.3  &  14210 &   0.23 &  ? &  $<-0.02$/$<0.66$    &K,W,SDSS(sp),S,LH,H \\
1331$-$0108     & 1.881 & 17.87 &  38.2 & $  0.59$ &  2.22 &   7911 & 50.3  &  18811 &   0.45 & Lo &   $ 0.37$/$0.64 $    &K,W,SDSS(sp),S,LH,H \\
\\								   
1442$-$0011     & 2.226 & 18.24 &  15.9 & $ -1.85$ &  3.58 &   5142 & 39.1  &  22834 &   0.39 & Hi &   $-0.20$/$<0.51$    &    K,W,SDSS,S,LH,H \\
1443$+$0141     & 2.451 & 18.20 &   9.2 & $ -2.36$ &  3.36 &   7967 & 44.0  &$>25000$ &  0.36 &   ? & $< 0.07$/$<0.77$    &     K,W,SDSS,S,LH,H \\
2111$-$4335     & 1.708 & 16.68 &  66.4 & $ -1.25$ &  4.19 &   7249 & 50.2  &  15322 &   0.51 & Hi &  $\cdots$/$\cdots$	 &            LBQS \\
2116$-$4439     & 1.480 & 17.68 &  41.0 & $ -1.90$ &  3.83 &   2594 & 30.2  &  24906 &   0.25 & Hi &  $\cdots$/$\cdots$   &          LH,LBQS\\
2140$-$4552     & 1.688 & 18.30 &  20.7 & $ -1.55$ &  2.54 &   1410 & 31.8  &  18944 &   0.23 & Hi &  $\cdots$/$\cdots$   &            LBQS \\
\\								   
2154$-$2005     & 2.035 & 18.12 &  15.1 & $ -1.90$ &  2.69 &    962 & 14.9  &  20019 &   0.00 & Hi &  $\cdots$/$\cdots$   &           K,W,LH,H \\
2201$-$1834     & 1.814 & 17.81 &  56.9 & $  0.47$ &  2.85 &   1612 & 27.3  &  19682 &   0.00 & Hi &  $<-0.45$/$\cdots$   &         K,W,S,LH,H \\
2211$-$1915     & 1.952 & 18.02 &  28.0 & $ -0.94$ &  2.30 &     27 & \phantom{1}7.8  &  11544 &   0.00 & Hi &  $\cdots$/2.11\tablenotemark{i}    &          W,LBQS    \\
2350$-$0045A    & 1.624 & 18.63 &  13.5 & $ -1.66$ &  3.12 &   6964 & 44.5  &  24718 &   0.57 & Lo &  $< 0.27$/$<0.67$    &K,W,SDSS(sp),S,LH,H \\
2358$+$0216     & 1.872 & 18.61 &  33.7 & $  0.95$ &  3.28 &   6283 & 48.6  &  23913 &   0.28 & Lo &  $\cdots$/$<0.62$\tablenotemark{i}   &               K,LH \\
\tableline
0010$-$0012\tablenotemark{k} & 2.145 & 18.46 & 7.3 & $-2.16$ & 3.07 & 0 &  2.8   & 14157 & 0.12 &$\cdots$ & $\cdots$/$\cdots$   & W,SDSS(sp),S\\
\enddata
\tablenotetext{a}{
The listed values of $z$ and \bj\ are taken from \citet{lbqs_ref}.
  $^{\rm b}$ The mean rest-frame $2500\pm25$\,\AA\ flux density
  ($\flam$) measured from spectra normalized to the SDSS (when
  available) or LBQS photometry.  $^{\rm c}$The spectral index of a
  power-law fit to the UV continuum where
  $f_{\lambda}\propto\lambda^{\alpha_{\rm UV}}$.  $^{\rm d}$The values
  for \nh\ (in units of 10$^{20}$\cmsq) are from Galactic \HI\ maps
  \citep{HI_ref}.  $^{\rm e}$The BALnicity Index (BI; as defined by
  \citealt{WeMoFoHe1991}) is a conservative measure of the \CIV\ \ew.
  Values from \citet{HewFol2003}.  $^{\rm f}$The maximum velocity of
  \CIV\ absorption blueshift (including the relativistic correction),
  \vmax. $^{\rm g}$The parameter \fdeep\ is the fraction of \CIV\
  absorption that absorbs $\ge50\%$ of the continuum. $^{\rm h}$Key:
  From the classification of \citet{LaHu2004}: Hi = UV spectra show
  only high ionization BALs; Lo = UV spectra also show either
  \ion{Al}{3} and/or \ion{Mg}{2} BALs; ?  = BAL type unknown because
  available UV spectra do not cover \ion{Mg}{2} region. $^{\rm
  i}$Radio-to-optical flux ratios, as defined in \S\ref{sec:sample}.
  Values marked with `$^{\rm i}$' use $S_{\rm 1.4GHz}$ rather than
  $S_{\rm 5GHz}$.  $^{\rm j}$References to available data on these
  quasars. Key: G = included in the \chandra\ survey of
  \citet{GreenEtal2001}; K = \citet{KoVoMoWe1993}; LBQS = LBQS
  discovery spectra were used to measure absorption-line and continuum
  parameters \citep{lbqs_ref}; LH = \citet{LaHu2004}; S =
  \citet{SirolaEtal1998}; SDSS(sp) = photometry available from the
  Sloan Digital Sky Survey (spectra available in Data Release 4); and
  W = \citet{WeMoFoHe1991}. $^{\rm k}$B0010$-$0012 has broad
  absorption evident in the LBQS and SDSS spectra, however, it does not meet
  the \balq\ criterion of \citet{WeMoFoHe1991} of BI$>0$.  Therefore,
  we present the X-ray data for the record, but do not include this
  object in the formal LBQS \balq\ sample.}
\end{deluxetable}

\clearpage
\end{landscape}
\clearpage
\begin{deluxetable}{lrlcrrrr}
\tabletypesize{\small}
\tablewidth{0pt}
\tablecaption{Observing Log
\label{tab:log}
}
\tablehead{
\colhead{Name\tablenotemark{a}} &
\colhead{Obs. ID} &
\colhead{Date} &
\colhead{Exposure} &
\multicolumn{2}{c}{Counts\tablenotemark{b}} &
\colhead{Count Rate\tablenotemark{b}} &
\colhead{\HR\tablenotemark{c}} \\
\colhead{(LBQS B)} &
\colhead{} &
\colhead{} &
\colhead{Time (ks)} &
\colhead{Soft} &
\colhead{Hard} &
\colhead{(10$^{-3}$\,ct\,\persec)} &
\colhead{} 
}
\startdata
0004$+$0147(Lo) &    4828  & 2003 Nov 09 & 5.54 &           $<2.3   $     &          $<2.3   $     &          $<0.42  $      &    $\cdots$         \\
0019$+$0107     &    2098  & 2001 Dec 08 & 4.80 &     7$^{+3.8 }_{ -2.6}$ &    6$^{+3.6 }_{ -2.4}$ &  2.50$^{+0.95}_{-0.71}$ & $ -0.08^{+0.33}_{-0.32}$\\
0021$-$0213     &    2099  & 2001 Aug 20 & 6.75 &     3$^{+2.9 }_{ -1.6}$ &          $<3.9   $     &  0.59$^{+0.47}_{-0.28}$ &          $<0.13  $     \\
0025$-$0151     &    2100  & 2001 Oct 31 & 4.69 &           $<5.5   $     &    4$^{+3.2 }_{ -1.9}$ &  1.28$^{+0.77}_{-0.51}$ &          $>-0.16 $     \\
0029$+$0017     &    2101  & 2001 Jun 20 & 6.69 &    32$^{+6.7 }_{ -5.6}$ &   10$^{+4.3 }_{ -3.1}$ &  6.28$^{+1.13}_{-0.97}$ & $ -0.52^{+0.16}_{-0.14}$\\
\\
0051$-$0019     &    4830  & 2004 Aug 05 & 7.11 &    18$^{+5.3 }_{ -4.2}$ &    9$^{+4.1 }_{ -2.9}$ &  3.80$^{+0.88}_{-0.73}$ & $ -0.33^{+0.22}_{-0.20}$\\
0054$+$0200     &    4831  & 2004 Aug 17 & 5.88 &     8$^{+4.0 }_{ -2.8}$ &    7$^{+3.8 }_{ -2.6}$ &  2.55$^{+0.84}_{-0.65}$ & $ -0.07^{+0.30}_{-0.29}$\\
0059$-$2735(Lo) &     813  & 2000 May 15 & 4.39 &           $<3.8   $     &          $<2.3   $     &          $<0.84  $      &    $\cdots$         \\
0109$-$0128     &    4832  & 2004 Nov 04 & 5.88 &           $<5.4   $     &    7$^{+3.8 }_{ -2.6}$ &  1.53$^{+0.70}_{-0.50}$ &          $>0.13  $     \\
1029$-$0125     &    2102  & 2001 May 30 & 4.50 &     7$^{+3.8 }_{ -2.6}$ &    4$^{+3.2 }_{ -1.9}$ &  2.44$^{+0.98}_{-0.72}$ & $ -0.27^{+0.36}_{-0.33}$\\
\\
1133$+$0214     &    4833  & 2004 Feb 13 & 5.91 &    12$^{+4.6 }_{ -3.4}$ &          $<5.2   $     &  2.37$^{+0.82}_{-0.63}$ &          $<-0.40 $     \\
1203$+$1530     &    4834  & 2005 Feb 05 & 7.06 &     3$^{+2.9 }_{ -1.6}$ &          $<3.8   $     &  0.57$^{+0.45}_{-0.27}$ &          $<0.12  $     \\
1205$+$1436     &    2103  & 2002 Mar 08 & 5.69 &    62$^{+8.9 }_{ -7.9}$ &   17$^{+5.2 }_{ -4.1}$ & 13.89$^{+1.75}_{-1.56}$ & $ -0.57^{+0.11}_{-0.10}$\\
1208$+$1535     &    2104  & 2002 Mar 08 & 4.90 &     4$^{+3.2 }_{ -1.9}$ &    2$^{+2.7 }_{ -1.3}$ &  1.23$^{+0.73}_{-0.48}$ & $ -0.33^{+0.51}_{-0.43}$\\
1212$+$1445     &    2456  & 2001 May 30 & 4.50 &           $<3.7   $     &          $<2.3   $     &          $<0.79  $      &    $\cdots$         \\
\\
1216$+$1103     &    2106  & 2001 Mar 18 & 4.64 &           $<3.9   $     &    5$^{+3.4 }_{ -2.2}$ &  1.29$^{+0.78}_{-0.51}$ &          $>0.12  $     \\
1230$+$1705     &    4835  & 2004 Feb 04 & 6.10 &     6$^{+3.6 }_{ -2.4}$ &          $<5.3   $     &  1.31$^{+0.65}_{-0.45}$ &          $<-0.06 $     \\
1231$+$1320(Lo) &    2107  & 2001 Aug 07 & 6.71 &           $<3.8   $     &          $<2.3   $     &          $<0.53  $      &    $\cdots$         \\
1235$+$0857     &    2108  & 2001 Mar 18 & 5.63 &    60$^{+8.8 }_{ -7.7}$ &   22$^{+5.8 }_{ -4.7}$ & 14.57$^{+1.79}_{-1.61}$ & $ -0.46^{+0.11}_{-0.10}$\\
1235$+$1453     &    2972  & 2002 Nov 17 & 6.65 &     2$^{+2.7 }_{ -1.3}$ &          $<5.3   $     &  0.60$^{+0.48}_{-0.29}$ &          $<0.45  $     \\
\\
1239$+$0955     &    2109  & 2001 Jul 25 & 5.16 &    10$^{+4.3 }_{ -3.1}$ &    7$^{+3.8 }_{ -2.6}$ &  3.29$^{+1.01}_{-0.79}$ & $ -0.18^{+0.28}_{-0.27}$\\
1240$+$1607     &    2973  & 2002 Nov 16 & 6.65 &     4$^{+3.2 }_{ -1.9}$ &    3$^{+2.9 }_{ -1.6}$ &  1.05$^{+0.57}_{-0.39}$ & $ -0.14^{+0.47}_{-0.44}$\\
1243$+$0121     &    2974  & 2002 May 03 & 6.67 &     6$^{+3.6 }_{ -2.4}$ &    4$^{+3.2 }_{ -1.9}$ &  1.50$^{+0.64}_{-0.47}$ & $ -0.20^{+0.38}_{-0.35}$\\
1314$+$0116     &    2975  & 2002 Jul 21 & 6.57 &           $<5.3   $     &    2$^{+2.7 }_{ -1.3}$ &  0.61$^{+0.48}_{-0.29}$ &          $>-0.45 $     \\
1331$-$0108(Lo) &    2110  & 2001 Mar 18 & 5.06 &     3$^{+2.9 }_{ -1.6}$ &    3$^{+2.9 }_{ -1.6}$ &  1.19$^{+0.71}_{-0.47}$ & $  0.00^{+0.50}_{-0.50}$\\
\\
1442$-$0011     &    2111  & 2001 May 30 & 4.28 &           $<3.8   $     &          $<2.3   $     &          $<0.86  $      &    $\cdots$         \\
1443$+$0141     &    2112  & 2001 Mar 23 & 5.94 &     6$^{+3.6 }_{ -2.4}$ &          $<3.9   $     &  1.18$^{+0.64}_{-0.43}$ &          $<-0.21 $     \\
2111$-$4335     &    4838  & 2003 Dec 16 & 5.53 &    15$^{+5.0 }_{ -3.8}$ &   17$^{+5.2 }_{ -4.1}$ &  5.79$^{+1.22}_{-1.02}$ & $  0.06^{+0.20}_{-0.20}$\\
2116$-$4439     &    4839  & 2004 Jun 21 & 5.65 &           $<2.3   $     &    3$^{+2.9 }_{ -1.6}$ &  0.53$^{+0.52}_{-0.29}$ &          $>0.13  $     \\
2140$-$4552     &    4840  & 2004 Mar 25 & 5.68 &     9$^{+4.1 }_{ -2.9}$ &    4$^{+3.2 }_{ -1.9}$ &  2.29$^{+0.83}_{-0.63}$ & $ -0.38^{+0.33}_{-0.28}$\\
\\
2154$-$2005     &    2113  & 2001 Sep 04 & 5.00 &     9$^{+4.1 }_{ -2.9}$ &    9$^{+4.1 }_{ -2.9}$ &  3.60$^{+1.07}_{-0.84}$ & $  0.00^{+0.27}_{-0.27}$\\
2201$-$1834     &    2114  & 2001 Aug 03 & 5.08 &           $<3.9   $     &          $<3.9   $     &          $<1.08  $      &    $\cdots$         \\
2211$-$1915     &    4836  & 2003 Nov 19 & 5.89 &    39$^{+7.3 }_{ -6.2}$ &   16$^{+5.1 }_{ -4.0}$ &  9.34$^{+1.44}_{-1.26}$ & $ -0.42^{+0.14}_{-0.13}$\\
2350$-$0045A(Lo)&    2115  & 2002 May 16 & 5.78 &           $<2.3   $     &          $<2.3   $     &          $<0.40  $      &    $\cdots$         \\
2358$+$0216(Lo) &    4837  & 2004 Jul 18 & 5.79 &           $<3.9   $     &          $<3.6   $     &          $<0.88  $      &    $\cdots$         \\
\tableline	     		  
0010$-$0012   &  4829  & 2004 Jun 27 & 6.66 &    13$^{+4.7 }_{ -3.6}$ &    3$^{+2.9 }_{ -1.6}$ &  2.40$^{+0.76}_{-0.59}$ & $ -0.63^{+0.27}_{-0.20}$\\
\enddata
\tablenotetext{a}{Known Lo\balqs\ are indicated with `(Lo)' following
  their names.}
\tablenotetext{b}{Detections for the full, soft, and hard bands are
  determined by {\em wavdetect}, and the counts are determined from aperture
  photometry (as described in \S\ref{sec:obs}). Errors are 1$\sigma$ Poisson errors \citep{Gehrels}, except for
  non-detections where the limits are the $90\%$ confidence limits
  from Bayesian statistics \citep*{kbn}. The count rate is for the full band, 0.5--8.0~keV.}
\tablenotetext{c}{The \HR\ is defined as $(h-s)/(h+s)$, where $h$ and
  $s$ are the counts in the hard (2.0--8.0~keV) and soft (0.5--2.0~keV)
  bands, respectively. The \HR\ errors are propagated from the
  counting errors using the numerical method of \citet{Lyons1991}.}
 \end{deluxetable}

\clearpage
\begin{landscape}
\begin{deluxetable}{lcccccrrrr}
\tablecolumns{10}
\tabletypesize{\scriptsize}
\tablewidth{0pt}
\tablecaption{X-ray Properties
\label{tab:xcalc}
}
\tablehead{
\colhead{Name\tablenotemark{a}} &
\colhead{\GHR\tablenotemark{b}} &
\colhead{$\log(F_{\rm X})$\tablenotemark{c}} &
\colhead{$\log(f_{\rm 2 keV})$\tablenotemark{d}} &
\colhead{$\log(f_{\rm 2500})$\tablenotemark{d}} &
\colhead{$\log(l_{\rm 2500})$\tablenotemark{e}} &
\colhead{\aox} &
\colhead{\daox\tablenotemark{f}} &
\colhead{\aoxc\tablenotemark{g}} &
\colhead{$\Delta$\aoxc\tablenotemark{h}} \\
\colhead{(LBQS B)} &
\colhead{} &
\colhead{} &
\colhead{} &
\colhead{} &
\colhead{} &
\colhead{} &
\colhead{} &
\colhead{} &
\colhead{}
}
\startdata
0004$+$0147(Lo) &    $\cdots$             & $<-14.350$        & $<-32.609$        & $-26.422$ &  31.440 & $<-2.37$ & $<-0.71$ & $<-1.96$ & $<-0.30$ \\
0019$+$0107     &  0.68$^{+0.59}_{-0.58}$ & $-13.498\pm0.143$ & $-32.001\pm0.143$ & $-26.406$ &  31.627 & $-2.15$ & $-0.46$  & $-1.76$ & $-0.07$ \\
0021$-$0213     &          $>0.32  $      & $-13.995\pm0.267$ & $-32.816\pm0.267$ & $-26.631$ &  31.458 & $-2.37$ & $-0.71$  & $<-1.79$ & $<-0.13$ \\
0025$-$0151     &          $<0.80  $      & $-13.960\pm0.213$ & $-32.369\pm0.213$ & $-26.420$ &  31.593 & $-2.28$ & $-0.60$  & $-1.82$ & $-0.14$ \\
0029$+$0017     &  1.52$^{+0.38}_{-0.34}$ & $-13.340\pm0.072$ & $-31.229\pm0.072$ & $-26.719$ &  31.357 & $-1.73$ & $-0.08$  & $-1.60$ & $ 0.05$ \\
\\
0051$-$0019     &  1.13$^{+0.42}_{-0.40}$ & $-13.441\pm0.091$ & $-31.611\pm0.091$ & $-26.636$ &  31.227 & $-1.91$ & $-0.28$  & $-1.69$ & $-0.06$ \\
0054$+$0200     &  0.67$^{+0.52}_{-0.53}$ & $-13.479\pm0.126$ & $-31.983\pm0.126$ & $-26.544$ &  31.389 & $-2.09$ & $-0.44$  & $-1.73$ & $-0.08$ \\
0059$-$2735(Lo) &    $\cdots$             & $<-14.053$        & $<-32.311$        & $-26.298$ &  31.507 & $<-2.31$ & $<-0.64$ & $<-1.97$ & $<-0.30$ \\
0109$-$0128     &          $<0.28  $      & $-13.487\pm0.168$ & $-32.285\pm0.168$ & $-26.567$ &  31.316 & $-2.19$ & $-0.55$  & $-1.72$ & $-0.08$ \\
1029$-$0125     &  1.02$^{+0.71}_{-0.64}$ & $-13.597\pm0.149$ & $-31.841\pm0.149$ & $-26.467$ &  31.529 & $-2.06$ & $-0.39$  & $-1.79$ & $-0.12$ \\
\\
1133$+$0214     &          $>1.28  $      & $-13.697\pm0.132$ & $-31.787\pm0.132$ & $-26.541$ &  31.198 & $-2.01$ & $-0.38$  & $<-1.80$ & $<-0.17$ \\
1203$+$1530     &          $>0.33  $      & $-14.020\pm0.266$ & $-32.762\pm0.266$ & $-26.774$ &  31.049 & $-2.30$ & $-0.69$  & $<-1.77$ & $<-0.16$ \\
1205$+$1436     &  1.65$^{+0.28}_{-0.25}$ & $-13.034\pm0.052$ & $-30.904\pm0.052$ & $-26.605$ &  31.225 & $-1.65$ & $-0.02$  & $-1.56$ & $ 0.07$ \\
1208$+$1535     &  1.13$^{+1.10}_{-0.91}$ & $-13.931\pm0.209$ & $-32.096\pm0.209$ & $-26.609$ &  31.360 & $-2.11$ & $-0.46$  & $-1.87$ & $-0.22$ \\
1212$+$1445     &    $\cdots$             & $<-14.076$        & $<-32.334$        & $-26.351$ &  31.471 & $<-2.30$ & $<-0.64$ & $<-1.96$ & $<-0.30$ \\
\\
1216$+$1103     &          $<0.30  $      & $-13.574\pm0.211$ & $-32.341\pm0.211$ & $-26.485$ &  31.334 & $-2.25$ & $-0.60$  & $-1.78$ & $-0.13$ \\
1230$+$1705     &          $>0.65  $      & $-13.758\pm0.179$ & $-32.249\pm0.179$ & $-26.718$ &  30.994 & $-2.12$ & $-0.52$  & $<-1.74$ & $<-0.14$ \\
1231$+$1320(Lo) &    $\cdots$             & $<-14.253$        & $<-32.511$        & $-26.207$ &  31.910 & $<-2.42$ & $<-0.70$ & $<-2.03$ & $<-0.31$ \\
1235$+$0857     &  1.40$^{+0.23}_{-0.23}$ & $-12.934\pm0.051$ & $-30.874\pm0.051$ & $-26.177$ &  32.085 & $-1.80$ & $-0.05$  & $-1.62$ & $ 0.13$ \\
1235$+$1453     &          $>-0.30 $      & $-13.818\pm0.271$ & $-33.264\pm0.271$ & $-26.829$ &  31.382 & $-2.47$ & $-0.82$  & $<-1.64$ & $<0.01 $ \\
\\
1239$+$0955     &  0.85$^{+0.53}_{-0.48}$ & $-13.419\pm0.117$ & $-31.789\pm0.117$ & $-26.454$ &  31.535 & $-2.05$ & $-0.38$  & $-1.73$ & $-0.06$ \\
1240$+$1607     &  0.78$^{+0.89}_{-0.84}$ & $-13.906\pm0.195$ & $-32.337\pm0.195$ & $-26.677$ &  31.434 & $-2.17$ & $-0.51$  & $-1.81$ & $-0.15$ \\
1243$+$0121     &  0.88$^{+0.72}_{-0.66}$ & $-13.779\pm0.159$ & $-32.135\pm0.159$ & $-26.363$ &  31.873 & $-2.22$ & $-0.50$  & $-1.86$ & $-0.14$ \\
1314$+$0116     &          $<1.30  $      & $-14.391\pm0.266$ & $-32.414\pm0.266$ & $-26.548$ &  31.659 & $-2.25$ & $-0.56$  & $-1.91$ & $-0.22$ \\
1331$-$0108(Lo) &  0.54$^{+0.94}_{-0.94}$ & $-13.764\pm0.211$ & $-32.366\pm0.211$ & $-26.179$ &  31.757 & $-2.37$ & $-0.67$  & $-1.98$ & $-0.28$ \\
\\
1442$-$0011     &    $\cdots$             & $<-14.031$        & $<-32.290$        & $-26.462$ &  31.604 & $<-2.24$ & $<-0.56$ & $<-1.87$ & $<-0.19$ \\
1443$+$0141     &          $>0.90  $      & $-13.868\pm0.193$ & $-32.206\pm0.193$ & $-26.643$ &  31.496 & $-2.13$ & $-0.46$  & $<-1.76$ & $<-0.09$ \\
2111$-$4335     &  0.43$^{+0.35}_{-0.35}$ & $-13.044\pm0.084$ & $-31.716\pm0.084$ & $-25.994$ &  31.867 & $-2.20$ & $-0.48$  & $-1.79$ & $-0.07$ \\
2116$-$4439     &          $<0.32  $      & $-14.006\pm0.321$ & $-32.744\pm0.321$ & $-26.279$ &  31.467 & $-2.48$ & $-0.82$  & $-1.99$ & $-0.33$ \\
2140$-$4552     &  1.23$^{+0.67}_{-0.59}$ & $-13.687\pm0.137$ & $-31.799\pm0.137$ & $-26.505$ &  31.346 & $-2.03$ & $-0.38$  & $-1.83$ & $-0.18$ \\
\\
2154$-$2005     &  0.54$^{+0.48}_{-0.48}$ & $-13.283\pm0.114$ & $-31.896\pm0.114$ & $-26.538$ &  31.460 & $-2.06$ & $-0.40$  & $-1.65$ & $ 0.01$ \\
2201$-$1834     &    $\cdots$             & $<-13.936$        & $<-32.195$        & $-26.027$ &  31.881 & $<-2.37$ & $<-0.65$ & $<-2.00$ & $<-0.28$ \\
2211$-$1915     &  1.32$^{+0.28}_{-0.28}$ & $-13.106\pm0.062$ & $-31.145\pm0.062$ & $-26.294$ &  31.672 & $-1.86$ & $-0.17$  & $-1.68$ & $ 0.01$ \\
2350$-$0045A(Lo)&    $\cdots$             & $<-14.348$        & $<-32.606$        & $-26.713$ &  31.108 & $<-2.26$ & $<-0.65$ & $<-1.85$ & $<-0.24$ \\
2358$+$0216(Lo) &    $\cdots$             & $<-14.025$        & $<-32.283$        & $-26.237$ &  31.696 & $<-2.32$ & $<-0.63$ & $<-1.95$ & $<-0.26$ \\
\tableline
0010$-$0012  &  1.80$^{+0.75}_{-0.62}$ & $-13.832\pm0.121$ & $-31.567\pm0.121$ & $-26.821$ &  31.218 & $ -1.82$  & $-0.19$  & $ -1.77$ & $ -0.14$\\
\enddata
\tablenotetext{a}{Known Lo\balqs\ are indicated with `(Lo)' following
  their names.}
\tablenotetext{b}{\GHR\ is a coarse measure of the hardness of the X-ray
  spectrum determined by comparing the observed \HR\ (see
  Table~\ref{tab:log}) to a simulated \HR\ that takes into account
  spatial and temporal variations in the instrument response (see
  $\S$\ref{sec:obs}).}
\tablenotetext{c}{The full-band X-ray flux, $F_{\rm X}$, has units of \flux\ 
  and is calculated by integrating the power-law spectrum with \GHR\
  and normalized by the full-band count rate from 0.5--8.0~keV.  The errors
  are derived from the 1$\sigma$ errors in the full-band count rate.
  For calculating upper limits to $F_{\rm X}$ and $f_{\rm 2 keV}$ from
  non-detections, \GHR\ was set to 1.0.}
\tablenotetext{d}{X-ray and optical flux densities were measured at
  rest-frame 2~keV and 2500\,\AA, respectively; units are \fnu.}
\tablenotetext{e}{The 2500\,\AA\ monochromatic luminosity, $l_{\rm
    2500}$, has units of \lumin~Hz$^{-1}$.  The redshift bandpass
  correction has been included.}
\tablenotetext{f}{The parameter \daox\ is the difference between the
  observed \aox\ and \aoxl, the predicted \aox\ from $l_{\rm 2500}$
  calculated from Equation 6 of \citet{Strateva+2005}.}
\tablenotetext{g}{The parameter \aoxc\  is \aox\ calculated
  assuming $\Gamma=2.0$ and using the hard-band count rate to
  normalize the X-ray continuum.}
\tablenotetext{h}{$\Delta$\aoxc=\aoxc--\aoxl.}
\end{deluxetable}

\clearpage
\begin{deluxetable}{lccccccc}
\tablecolumns{8}
\tablewidth{0pt}
\tablecaption{Distributions of \aox, \daox, and $\Delta$\aoxc\tablenotemark{a} 
\label{tab:aox}}
\tablehead{
\colhead{Percentile} &
\multicolumn{2}{c}{\aox} &
\multicolumn{3}{c}{\daox} &
\multicolumn{2}{c}{$\Delta$\aoxc} \\
\colhead{}           &
\colhead{All}        &
\colhead{HiBALs}     &
\colhead{All}        &
\colhead{HiBALs}     &
\colhead{SDSS\tablenotemark{b}}       &
\colhead{All}        &
\colhead{HiBALs}     
}
\startdata
25\%          & $-2.47$ & $-2.30$ & $-0.73$ & $-0.68$ & $-0.08$\phantom{$-$} & $-0.25$ & $-0.33$ \\
50\%          & $-2.20$ & $-2.15$ & $-0.54$ & $-0.48$ & $0.02$  & $-0.20$ & $-0.14$ \\
75\%          & $-2.08$ & $-2.05$ & $-0.40$ & $-0.39$ & $0.09$  & $-0.08$ & $-0.07$ \\
\tableline
Median        & $-2.20$ & $-2.12$ & $-0.52$ & $-0.46$ &	$0.02$  & $-0.14$ & $-0.13$  \\
Mean          & $-2.21\pm0.04$ & $-2.14\pm0.04$ & $-0.53\pm0.04$ &
$-0.48\pm0.04$ & $-0.01\pm0.02$ &  $-0.18\pm0.02$ & $-0.15\pm0.03$ \\
\enddata
\tablenotetext{a}{The numbers in the first three rows are the highest 
values in each quartile listed in the first column. The
  values in this table (except for the medians) were calculated using
  the Kaplan-Meier estimator with {\sc ASURV} Rev 1.2
  \citep{LaIsFe1992} implementing the methods of \citet{FeNe1985}. The
  column headings `All' and `HiBALs' refer to the entire sample of 35
  \balqs\ and the subsample of 24 known Hi\balqs, respectively (see
  Table~\ref{tab:opt}).}
\tablenotetext{b}{The `SDSS' column heading refers to the sample from
  \citet{Strateva+2005} presented in Figure~\ref{fig:hist}.}
\end{deluxetable}

\clearpage
\end{landscape}
\begin{deluxetable}{llcrc}
\tablecolumns{5}
\tablewidth{0pt}
\tablecaption{Results from Non-Parametric Bivariate Statistical Tests\tablenotemark{a}
\label{tab:stats}}
\tablehead{
\colhead{Variables\tablenotemark{b}} &
\multicolumn{2}{c}{Generalized Kendall} &
\multicolumn{2}{c}{Spearman} \\
\colhead{Independent/Dependent} &
\colhead{$\tau$}        &
\colhead{Prob.\tablenotemark{c}}        &
\colhead{$\rho$}        &
\colhead{Prob.\tablenotemark{c}} 
}
\startdata
\daox/\GHR\ (27)                     &   4.055 & 0.0001 & $\cdots$  & $\cdots$\\
\aox/\GHR\ (27)                      &   4.156 & $<0.0001$ & $\cdots$  & $\cdots$\\
BI/\daox\ (35)                       &   2.134 & 0.033  & $-0.350$  &  0.041 \\
BI/\daox\ (HiBALs only -- 24)        &   1.582 & 0.114  & $-0.332$  &  0.111 \\
DI/\daox\ (27)                       &   0.711 & 0.477  & $-0.159$  &  0.418 \\
DI/\daox\ (HiBALs only -- 17)        &   0.418 & 0.676  & $-0.116$  &  0.643 \\
\vmax/\daox\ (35)                    &   2.924 & 0.003  & $0.517$   &  0.003 \\
\vmax/\aox\ (35)                     &   2.876 & 0.004  & $0.517$   &  0.003 \\ 
\vmax/\daox\  (HiBALs only -- 24)    &   1.883 & 0.060  & $0.406$   &  0.052 \\
\fdeep/\daox\ (35)                   &   0.102 & 0.918  & $0.014$   &  0.935 \\
\fdeep/\daox\ (HiBALs only -- 24)    &   0.050 & 0.960  & $-0.001$  &  0.998 \\
\auv/\daox\ (35)                     &   1.213 & 0.225  & $-0.217$  &  0.206 \\
\auv/\daox\ (HiBALs only -- 24)      &   0.126 & 0.900  & $0.030$   &  0.887 \\
$l_{2500}$/\vmax\ (35)               &   0.284 & 0.776  & $-0.011$  &  0.951 \\
$l_{2500}$/\vmax\ (HiBALs only -- 24)               &   0.248 & 0.804  & $0.050$   &  0.809 \\
\CIV\ EW$_{\rm a}$/\daox\ (BQS -- 41) &   4.816 & $<0.0001$ & $-0.694$ & $<0.0001$ \\ 
\enddata
\tablenotetext{a}{Bivariate statistical tests were performed using
{\sc ASURV} Rev 1.2 \citep{LaIsFe1992} which implements the methods
described in \citet{IsFeNe1986}.}  \tablenotetext{b}{The number of
data points is given in parentheses.}  \tablenotetext{c}{The
probability that the given variables are not correlated. 
Spearman's $\rho$ cannot be calculated for data with both upper and
lower limits, such as \GHR.}
\end{deluxetable}

\end{document}